\documentclass[graybox]{svmult}

\usepackage{mathptmx}       
\usepackage{helvet}         
\usepackage{courier}        
\usepackage{type1cm}        
\usepackage{makeidx}         
\usepackage{graphicx}        
\usepackage{multicol}        
\usepackage[bottom]{footmisc}

\usepackage{amsmath, amssymb}

\makeindex             

\begin{document}

\title*{Data mining when each data point is a network}
\author{Karthikeyan Rajendran, Assimakis Kattis, Alexander Holiday, Risi Kondor, Ioannis G. Kevrekidis}
\institute{Karthikeyan Rajendran \at  Department of Chemical \& Biological Engineering, Princeton University\\ Present Address: 731 Lexington Avenue, 10022, New York, NY, USA, email: karthikeyan.rajendran@gmail.com
\and Assimakis Kattis \at Theory Group, Department of Computer Science, University of Toronto\\ 10 King's College Road, M5S 3G4, Toronto, ON, Canada, email: kattis@cs.toronto.edu \and Alexander Holiday \at Department of Chemical \& Biological Engineering, Princeton University\\ 41 Olden Street, 08544, Princeton, NJ, USA, email: holiday@princeton.edu \and Risi Kondor \at Machine Learning Group, Computer Science \& Statistics, University of Chicago\\ Ryerson 257B, 1100 E. 58th Street, 60637, Chicago, IL, USA, email: risi@cs.uchicago.edu \and Ioannis G. Kevrekidis \at A319 Engineering Quad, Department of Chemical \& Biological Engineering and Program in Applied \& Computational Mathematics, Princeton University\\ Technische Universit\"{a}t M\"{u}nchen - Institute for Advanced Study; Zuse Institut Berlin\\41 Olden Street, 08544, Princeton, NJ, USA, email: yannis@princeton.edu}
%
%
\maketitle

\noindent \textit{keywords:} networks, dimensionality reduction, equation free, dynamical systems, diffusion maps\\

\noindent \textit{MSC 2010 codes:} 37 Dynamical Systems and Ergodic Theory, 65 Numerical Methods, 68 Computer Science\\

\abstract{ We discuss the problem of extending data mining approaches to cases in which data points arise in the form of individual graphs. Being able to find the intrinsic low-dimensionality in ensembles of graphs can be useful in a variety of modeling contexts, especially when coarse-graining the detailed graph information is of interest. One of the main challenges in mining graph data is the definition of a suitable pairwise similarity metric in the space of graphs. We explore two practical solutions to solving this problem: one based on finding subgraph densities, and one using spectral information. The approach is illustrated on three test data sets (ensembles of graphs); two of these are obtained from standard graph generating algorithms, while the graphs in the third example are sampled as dynamic snapshots from an evolving network simulation. We further incorporate these approaches with equation free techniques, demonstrating how such data mining approaches can enhance scientific computation of network evolution dynamics.}

\section{\label{sec:intro} Introduction}

Microscopic modeling and simulation are increasingly being used
to describe complex systems, as they are easy to implement and often
describe physical systems with great accuracy. Microscopic modeling is extensively used in various
fields such as epidemiology \cite{Euba04modelling,Ferg05strategies,Long86predicting},
economics \cite{Iori02microsimulation,Wang05microscopic},
biology \cite{Levi01self-organization,Liu04stable}, and so on. Networks are identified as a key feature of the structure of many
such complex systems \cite{Bara02linked:,Newm03structure}. When the sizes of the networks become large, it is  important to find tools to systematically analyze the networks. In this paper, we will focus on the issue of data mining, and the challenges involved in extending standard data mining approaches to cases where every data point is a graph. We also discuss the efficient estimation of various system quantities using data mining techniques, which build significantly on previous work \cite{rajendran2013analysis}.

There are numerous applications in which data mining approaches on graphs
would be useful. They can be used to find the dimensionality of the
subspace in which any given collection of graphs lives. In other words, data mining algorithms applied to collections of graphs can help us understand the number of important variables required to characterize (and thus parametrize) them. Being able to decipher the minimum number of variables
required to represent graphs is useful in itself. One can then take the additional postprocessing step of finding the relationship between variables extracted from data mining and actual network properties. However, this {\em mapping} from data mining variables to actual variables is a separate self-contained problem. We first have to look at problems where there is enough understanding about the graph datasets, using this intuition to validate the results of our data mining procedure.

A separate class of problems in which data mining approaches can be of
crucial utility are those where the graph datasets come from a
dynamical process. In such cases, data mining can help us understand the dynamics
of the process. As before, in order to relate the data mining results to actual properties
of the system, one has to perform additional processing to map the data mining
variables to real system properties. We illustrate such methods on
sets of graphs created by different algorithmic processes in
their full parameter space. This allows us to obtain
the dimensionality of the space in which these graphs live i.e., to understand the actual variation in the graphs
produced by each of these algorithms, which is crucial in understanding whether all the parameters
in a model are independent. One can then seek ways to efficiently parameterize the graphs. To this end, we address the independent issue of efficiently mapping between the extracted variables and the underlying network properties, and use this to demonstrate how to accelerate the estimation of desired dynamical quantities from the underlying process. This approach to studying graph generating algorithms may also be used to propose and test more generalized algorithms for generating graphs that sample more of the space of all possible graphs with a given size and so on. Such an algorithm can find use in parametric optimization contexts in helping to construct graphs with prescribed collective properties \cite{Goun11generation}.

The rest of the paper is organized as follows: In Section \ref{sec:dm}, we briefly discuss the data mining algorithm
that will be used in this paper. In Section \ref{sec:sim}, we focus on the issue of defining similarities between graphs,
which is the biggest challenge in adapting traditional data mining techniques to this context. In the same section, we also discuss two options for solving this problem. We take three illustrative examples and implement the data mining algorithm with our two choices of similarity measures in Section ~\ref{sec:res}. In Section ~\ref{sec: ef}, we discuss the mapping between the obtained data mining variables and underlying properties of the network, also providing an efficient method for the estimation of underlying quantities of interest from a network-based dynamical system. A summary of results and suggestions for future work are presented in Section ~\ref{sec:conc}.

\section{\label{sec:dm}Data mining}

The most traditional tool in data mining is principal component analysis (PCA) \cite{Shle03tutorial},
which is used to represent a low dimensional dataset, represented in high dimensional space,
in terms of the most meaningful linear basis. It enables one to identify directions in which the data points have the most variance. But PCA is only a linear analysis tool as it can only find out the best `linear' lower
dimensional subspace in which the dataset lives. In many problems, the data lives in a highly \textit{non-linear}
lower dimensional subspace making the low dimensional \textit{linear} subspace much higher dimensional
compared to the true dimensionality of the space in which the data lie. A number of non-linear data mining tools such as Diffusion maps \cite{Nadl05diffusion,Nadl06diffusion} and ISOMAPs \cite{Tene00global} are available to extract the non-linear subspace. In this work, we use diffusion maps as a representative non-linear data mining approach
in order to enable our discussion on extending these approaches to datasets represented
in the form of graphs. In Diffusion maps (DMAPs), one constructs a graph with the data points as vertices; a similarity measure between the data points is used as weights on the edges. In broad terms, the eigenfunctions of the diffusion process on this graph are used to embed the data points. If the data points actually live in a low dimensional non-linear subspace, the first few of these eigenfunctions will be enough to embed the data and
still be able to recover all the information about the data. A brief discussion of diffusion maps is given below.

Consider a set of $n$ points ${x_i}_{i=1}^n \in \mathbb{R}^p$. We define a similarity matrix, $W$ (which is a measure of closeness between pairs of points in this space) according to the following equation:

\begin{equation}
W(i,j) = exp \left( \frac{-\|x_i-x_j\|^2}{\epsilon^2} \right).
\label{eq:sim}
\end{equation}

This choice of similarity matrix is called a Gaussian kernel. Here, $\epsilon$ is a suitable length scale characterizing the neighborhood of the point. Let us also define a diagonal normalization matrix, $D_{ii}=\sum_j W_{ij}$, and consequently the matrix, $A=D^{-1} W$. $A$ can be viewed as a Markov matrix defining a random walk (or diffusion on the data points), i.e., $A_{ij}$ denotes the probability of transition from  $x_i$ to $x_j$. Since $A$ is a Markov matrix, the first eigenvalue is always $1$. The corresponding eigenvector is a constant, trivial eigenvector. In diffusion maps, the next few non-trivial eigenvectors of $A$
(corresponding to the next few largest eigenvalues) are the best directions
that span the non-linear subspace in which the data lives. Thus, these eigenvectors are used to characterize this non-linear manifold.

As is evident in the description above, an important step in the implementation
of diffusion maps is the definition of a measure of similarity between data points.
If the data points live in a Euclidean space, it is straightforward to use the
Euclidean distance to measure the distance (or the closeness) between pairs of points (or graphs).
When the data points are represented in the form of graphs, however, it is not
trivial to define good measures of similarity between them.
Thus, if all the machinery of non-linear data mining is to be successfully
adapted to the case of graph data, one has to be able to define a measure
of similarity and closeness between pairs of graphs.

\section{\label{sec:sim} Defining similarity measures between graph objects}

Although measures of similarity in the context of graphs have been discussed in the
literature \cite{Dana11algorithms}, complete systematic classifications
and definitions are still lacking.
Firstly, one can either define similarities between nodes in a given graph
or similarities between graphs themselves.
In this paper, we will discuss the latter type,
since we are interested in comparing entire graph objects.
Secondly, the nodes of the graphs can be labeled or unlabeled.
We are interested in the case of unlabeled nodes,
where the problem of ordering the nodes makes it more challenging
to define similarity measures.
Additionally, we will focus on the case where all the graphs in the dataset have
the same number of nodes.
However, the approach is, in principle, extendable to collections of
graphs of different sizes.

Existing techniques in the literature for defining similarities
may roughly be classified into a few broad categories.
The first of these is the class of methods that make use of
the {\em structure} of the graphs to define similarities.
An obvious choice is to consider two graphs to be similar
if they are isomorphic \cite{Peli98replicator}.
One of the first definitions of distance between pairs of graphs
using the idea of graph isomorphism was based on constructing the
smallest larger graph whose subset was isomorphic to both the graphs \cite{Zeli75certain}.
Likewise, one can define similarity measures based on the largest
common subgraph in pairs of graphs \cite{Bunk98graph,Raym02rascal:}.
The graph edit distance, which measures the number of operations on the nodes
and edges of the graph required to transform one graph into another, is another
example of a method using the idea of graph isomorphism.
The graph edit distance and a list of other measures that use the
structure of the network to quantify similarity are defined in \cite{Papa08web}.

Next we have iterative methods that compare the behavior of the neighborhoods of
the nodes in the graphs.
Comparing neighborhoods of nodes is especially applicable to measure similarities
between sparse graphs.
Often, the graph similarity problem is solved through solving the related
problem of graph matching, which entails finding the correspondence between
the nodes in the two graph such that the edge overlap is maximal.
Methods like the similarity flooding algorithm \cite{Meln02similarity}, the graph similarity scoring algorithm \cite{Zage08graph} and the belief
propagation algorithm \cite{Bay09message} are a few such approaches.
Graph kernels based on the idea of random walks \cite{Gaert03graph,Kash03marginalized,Mahe04extensions}
also fall under the category of algorithms based on comparing neighborhoods.

However, one of the simplest options to evaluate similarities between graphs is to
directly compare a few chosen, representative features of the network.
The chosen features may correspond to any facet of the graph, such as
structural information (degree distribution, for instance) or spectral measures
(eigenvalues and/or eigenvectors of the graph Laplacian matrix).
In this paper, we will take this approach and consider two options
for defining similarities between graphs.
The two options for defining similarity measures between graphs considered here are:
(i) using subgraph densities and (ii) an approach using spectral information.
A detailed description of the two proposed measures of graph similarity follows.

\subsection{\label{ss:den} Subgraph density approach}

The general idea behind this approach is that two graphs
are similar if {\em the frequency of occurrence of representative subgraphs
in these graphs are similar.}
Density of a small subgraph in a large graph is a weighted
frequency of occurrence of the subgraph (pattern) in the large, original graph.
We use the following definition for the subgraph density of a subgraph $H$ with $k$ nodes
in a graph $G$ with $n$ nodes:

\begin{align}
\label{eqn:homdenG}
\rho(H,G) := \frac{1}{{n\choose k}} \!  \sum_{\varphi:[k] \to [n]} \!
\left[ \forall  i, j \in [k] \! : \! H(i,j) \! = \! G_n(\varphi(i),
\varphi(j)) \right].
\end{align}

A graph can be reconstructed exactly if the densities of all possible
subgraphs onto the graph are specified \cite{Lov06limits}.
Thus, a list of all these subgraph densities is an alternative way
to provide complete information about a graph.
This list can be thought of as an embedding of the graph, which can
then be used to define similarity measures in the space of graphs.
It is, however, not practical to find the subgraph densities of all
possible subgraphs of a given graph, especially when the number of
nodes in the graph becomes large.
A systematic, yet practical, way to embed a graph is to use these subgraph densities
of all subgraphs lesser than a given size onto the graph.
For instance, to embed a graph with $n$ nodes, one can evaluate the subgraph densities
of all subgraphs of size less than or equal to $m~(m<<n)$.
Since $m<<n$, the embedding cannot be used to exactly reconstruct the graph.
One can, nevertheless, compute distances between the embeddings
(the vector of subgraph densities) of any two graphs and use them
to estimate similarities between these graphs.

Let $G_i$ and $G_j$ be two graphs defined on $n$ nodes.
Let $H_1$, $H_2$,$\ldots H_r$ be the $r$ chosen, representative subgraphs.
We find the frequencies of occurrence of these subgraphs in the original
graphs by appropriately modifying the open-source RANDESU algorithm described
in \cite{Wern06fanmod:}.
The subgraph densities are calculated by dividing these frequencies by
$n\choose{k}$, where $n$ and $k$ are the the number of nodes in the
original graph and the subgraph respectively.
(Note that although dividing by $n\choose{k}$ is not a unique choice
for normalizing the subgraph densities, the densities we calculated this way had similar
orders of magnitude, and hence, this constitutes a sensible choice.)
The density of subgraph $H$ in graph $G$, denoted by $\rho(H,G))$, is calculated
as mentioned above.
The similarity measure between a pair of graphs $G_i$ and $G_j$ can then
be defined as an $L_2$-norm (possibly weighted) of the difference between the vector of subgraph densities
as follows:

\begin{equation}
k(G_i,G_j) = \sqrt{\sum_{l=1}^{r}\left(\rho(H_l,G_i)-\rho(H_l,G_j)\right)^2}.
\label{eq:k1}
\end{equation}

In order to use this pairwise similarity measure in a diffusion map context, the
Gaussian kernel, analogous to Eq.~\ref{eq:sim}, can be calculated as follows:

\begin{equation}
W(i,j) = exp \left( \frac{-(k(G_i,G_j))^2}{\epsilon^2} \right).
\label{eq:sim1}
\end{equation}

\begin{figure}
\begin{center}
\includegraphics[width=0.63\textwidth]{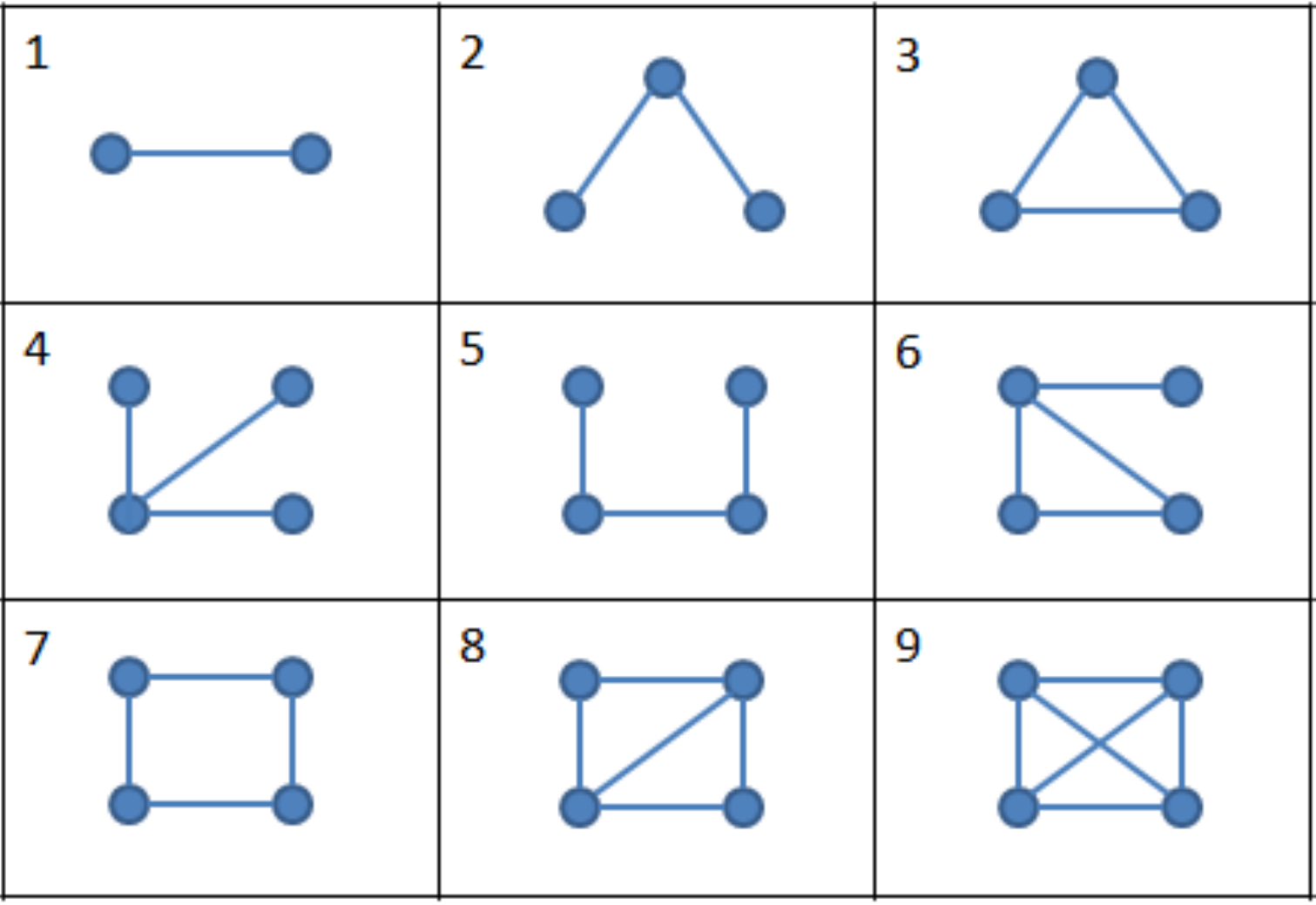}
\caption{\label{fig:SGs} The $9$ connected subgraphs of size less than or equal to $4$.
}
\end{center}
\end{figure}

In our illustrative numerical computations, we considered all connected subgraphs of size less than
or equal to $m=4$ as a representative sample of subgraphs.
There are $r=9$ such graphs as shown in Fig.~\ref{fig:SGs}.

\subsection{\label{ss:s}Spectral approach}

Our second approach to defining similarities between graphs was initially motivated by the approach given
in \cite{Vis10graph}, and we based it on the notion of non-conservative diffusion on graphs \cite{Gho11non-conservative}.
It has to be noted here that there are numerous ways in which the spectral
information of graphs (or equivalently information from performing random walks on graphs)
could be used to define similarity measures.
The particular version of the similarity metric discussed here is inspired by the
spectral decomposition algorithm in \cite{Vis10graph}.
The usual definition of random walks on graphs is based on the physical diffusion process.
One starts with a given initial density of random walkers, who are then redistributed at
every step by premultiplying the distribution of random walkers at the current stage by
the adjacency matrix.
The rows of the adjacency matrix are scaled by the row sum
so that the quantity of random walkers is conserved.
In our approach, we consider a non-conservative diffusion process where we replace the
normalized adjacency matrix in the random walk process by its original, unnormalized counterpart.

Let us consider two graphs $G_i$ and $G_j$, with adjacency matrices $B_i$ and $B_j$ respectively.
Let their spectral decompositions be given by $B_i = P_i D_i P_i^{T}$ and $B_j = P_j D_j P_j^{T}$ respectively.
Let the initial probability distribution of random walkers on the $n$ nodes of the graph be denoted by $\hat{p}$.
This can be taken to be a uniform distribution.
At every step of the process, the new distribution of random walkers is found
by applying the unnormalized adjacency matrix to the distribution at the previous step.
Since the adjacency matrix is not normalized, the density of random walkers
change over time depending on the weights associated with the edges of the graphs.
We consider walks of different lengths, at the end of which we evaluate
statistics by weighing the density of random walkers on the nodes according
to vector $\hat{q}$, which can also be assumed to be a uniform vector
that takes the value $1/n$ corresponding to every node.
As pointed out in \cite{Vis10graph}, the vectors $\hat{p}$ and $\hat{q}$
are ways to \emph{``embed prior knowledge into the kernel design"}.
Although the method is general, we will consider the special case
where the sizes of the graphs are the same.

The (possibly weighted) average density of random walkers after a k-length walk in $G_i$, denoted by $Q_{ik}$.
This can be evaluated as follows:

\begin{equation}
Q_{ik} = \hat{q}^\mathrm{T}B_i^k \hat{p} = \hat{q}^\mathrm{T} (P_i D_i^k P_i^{T}) \hat{p}.
\end{equation}

Consider a summation of $Q_{ik}$ for walks of all lengths with appropriate weights $\mu_k$
corresponding to each value of $k$.
Where $l_i=P_i^\mathrm{T}\hat{q}$ and $r_i=P_i^{T} \hat{p}$, let the computed weighted sum of densities corresponding to graph $G_i$ be denoted as $S_i$:

\begin{equation}
S_i = \sum_{k=0}^{\infty} \mu(k) Q_{ik} = \sum_{k=0}^{\infty} \mu(k)~l_i^\mathrm{T} D_i^k r_i.
\end{equation}

We used the following choice of weighting relation: $\mu(k) = \frac{\lambda^k}{k!}$.
With this choice of weights, one can write $S_i$ as a simple function of $\lambda$ as follows:

\begin{equation}
S_i(\lambda) = l_i^\mathrm{T}e^{(\lambda D_i)}r_i.
\label{eq:S}
\end{equation}

Thus, every graph $G_i$ is embedded using these $S_i$ values evaluated at characteristic
values of $\lambda$ (say $\lambda_1, \lambda_2,... \lambda_M$)\footnote{Note that
an alternative equivalent way to define the similarity measure would be to directly
compare the contribution of the different eigenvectors to $S_i$ instead of
summing the contributions and then using different values of $\lambda$.
However, it is difficult to generalize this approach
to cases where there are graphs of varying sizes.}.
The similarity between any two graphs $G_i$ and $G_j$ can then be evaluated using the
Gaussian kernel defined in Eq.~\ref{eq:sim1} using the following expression for $k(G_i,G_j)$:

\begin{equation}
k(G_i,G_j) = \sqrt{\sum_{m=1}^{M} (S_i(\lambda_m)-S_j(\lambda_m))^2}.
\label{eq:k2}
\end{equation}

This formula is very convenient for our purpose.
For every graph $G_i$, one can evaluate the three vectors,
$l_i$, diagonal elements of $D_i$ and $r_i$ and store them.
These $3n$ numbers can be thought of as a coarse embedding of the graph.
The similarity measure between pairs of graphs can finally be
evaluated by using Eq.~\ref{eq:S} and Eq.~\ref{eq:k2}
by substituting in these stored values.
This also makes it easier to add new graphs and increase the size of
the similarity matrix without having to do too much additional computation.

\section{\label{sec:res} Computational results}

We will explore the dimensionality of datasets (where the data points are individual
graphs) using the diffusion map approach; within this approach we will construct
implementations using the graph similarity metrics mentioned above.
We use three different datasets for this exploration; two of them arise in the
context of ``graph-generation" models (they are the ubiquitous Erd\"{o}s-R\'{e}nyi
networks and the Chung-Lu networks).
The third is closer to the types of applications
that motivated our work: networks that arise as individual temporal ``snapshots" during a
dynamic network evolution problem.
%
%

\subsection{\label{ss:er} Test case $1$: Erd\"{o}s-R\'{e}nyi graphs}

\begin{figure}
\begin{center}
\includegraphics[width=0.81\textwidth]{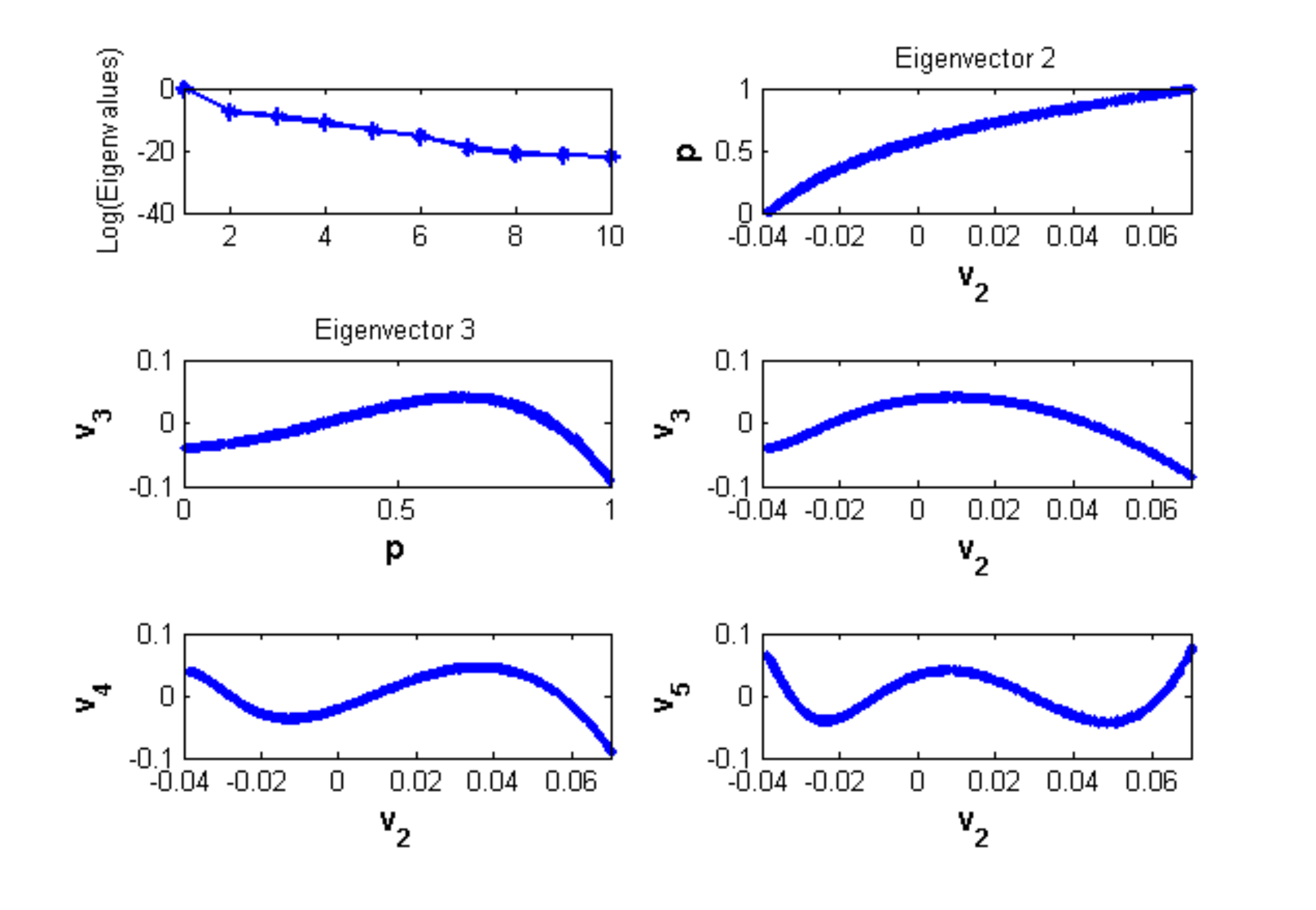}
\caption{\label{fig:ER1} Data mining ensembles of Erd\"{o}s-R\'{e}nyi graphs: The
subgraph approach was used to quantify similarity between individual graphs (see text).
The top-left plot shows the first $10$ eigenvalues of the random walk matrix arising in Diffusion Maps.
The corresponding first two non-trivial eigenvectors are plotted against the ``construction parameter" $p$
used to create the graphs, as well as against each other.
Notice how the first non-trivial eigenvector (the second eigenvector) is one-to-one with $p$.
}
\end{center}
\end{figure}

\begin{figure}
\begin{center}
\includegraphics[width=0.81\textwidth]{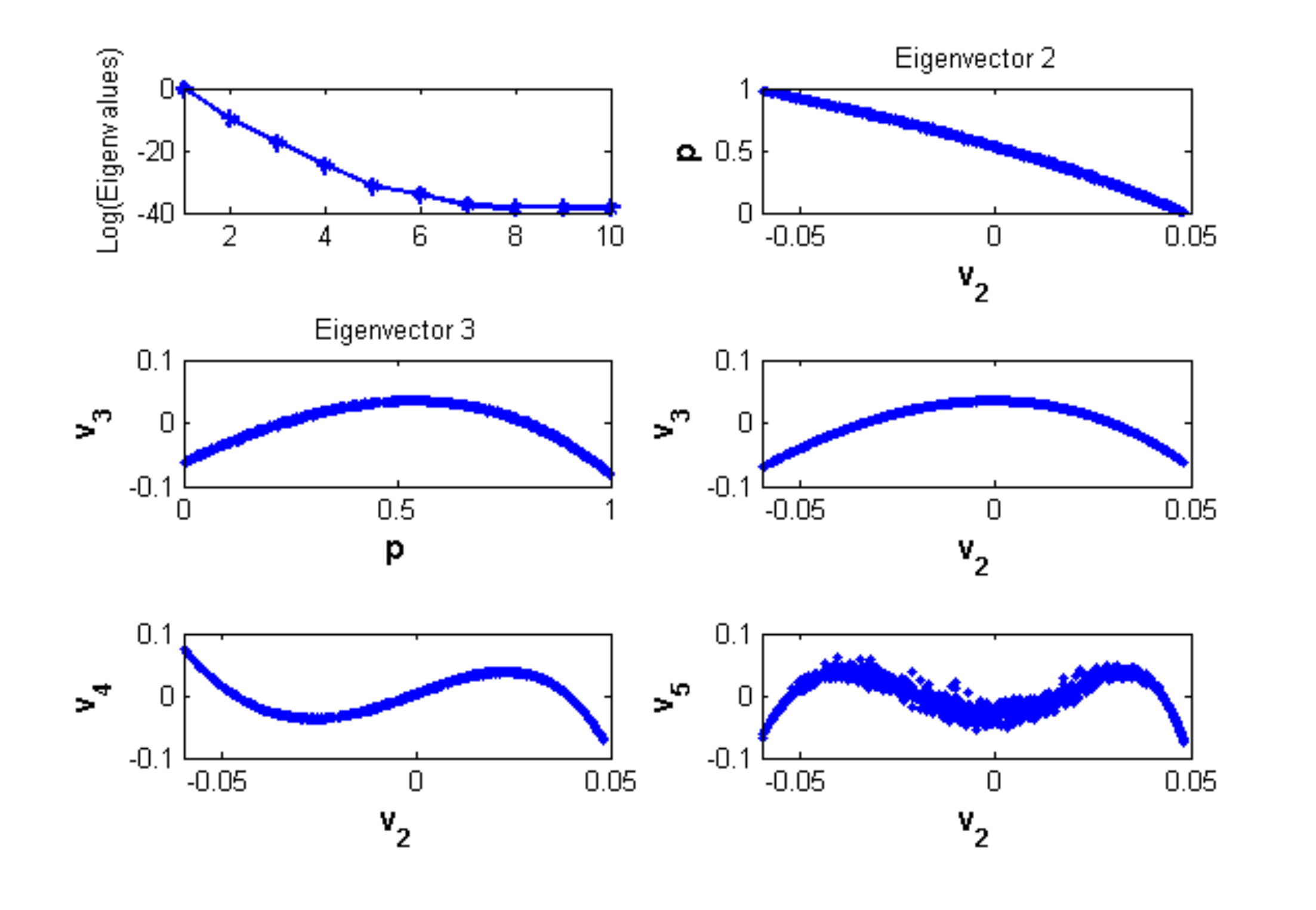}
\caption{\label{fig:ER2} Data mining ensembles of Erd\"{o}s-R\'{e}nyi graphs: Our
spectral approach was used to quantify similarity between graphs (see text). The top-left plot shows
the first $10$ eigenvalues of the random walk matrix arising in Diffusion Maps. The
corresponding first two non-trivial eigenvectors are plotted against the construction parameter $p$
used to create the graphs, as well as against each other.
Notice how, again, the first non-trivial eigenvector (the second eigenvector) is
one-to-one with $p$.
}
\end{center}
\end{figure}

Consider, as our initial example, a dataset consisting of $m=1000$
Erd\"{o}s-R\'{e}nyi $G(n,p)$ random graphs \cite{Erd59random} with $n=100$ nodes each.
The parameter $p$ (the probability of edge existence) used to construct these
graphs is randomly sampled uniformly in the interval $(0,1)$.
The diffusion maps algorithm is then applied on this set of graphs.
We start by computing the similarity measures between pairs of individual graphs
(both the subgraph (Eq.~\ref{eq:k1}) -using $9$ subgraph densities-
and our spectral approach (Eq.~\ref{eq:k2} -using 100 $\lambda$ values-).
The similarity matrix $W$ is then calculated using Eq.~\ref{eq:sim1}.
The first $10$ eigenvalues of the corresponding random walk matrix $A$,
(as described in Sec.~\ref{sec:dm}) are plotted in Figs.~\ref{fig:ER1} and \ref{fig:ER2},
corresponding to the subgraph and to our spectral approach respectively.
For both these cases, the first two non-trivial eigenvectors (viz., the eigenvectors corresponding
to the second and third
eigenvalues) are plotted against the parameter $p$ of the corresponding Erd\"{o}s-R\'{e}nyi graph.
From the figures, it is clear that the second eigenvector is
one-to-one with the parameter $p$, which here is also the edge-density.
Thus, this eigenvector (in both cases) captures the principal direction
of variation in the collection of Erd\"{o}s-R\'{e}nyi graphs.
In other words, our data mining approach independently recovers the
single important parameter $p$ in our sample dataset.

As these Erd\"{o}s-R\'{e}nyi graphs can be parameterized using just a single parameter $p$,
one might expect a gap in the eigenspectrum after the second (first nontrivial) eigenvalue,
and also expect the remaining eigenvalues/vectors to correspond to some sort of ``noise": the variability
of sampling among Erd\"{o}s-R\'{e}nyi graphs of the same $p$.
Interestingly, no such eigengap can be observed in our plots after the second eigenvalue.
If, however, subsequent eigenvectors are plotted against the second one on our data, we
clearly observed that they are simply higher harmonics in its ``direction".
The third, fourth and fifth eigenvectors, in both cases, are clearly seen to be a non-monotonic function of $v_2$($p$)
but with an increasing number of ``spatial" oscillations, reminiscent of Sturm-Liouville type problem eigenfunction shapes.
These eigenvectors do not, therefore, capture new directions in the space of our sample graphs.
%

This simple example serves to illustrate the purpose of using
data mining algorithms on graph data.
In this case, we created a one parameter family of graphs,
characterized by the parameter $p$.
Using only the resulting graph objects, our data mining
approach successfully recovered a characterization of these graphs
equivalent to (one-to-one with) this parameter $p$.
One feature of this one-to-one correspondence between $p$ and the $v_2$ component
of the graphs is worth more discussion: data mining discovers the ``one-dimensionality"
of the data ensemble, but does not {\em explicitly} identify $p$ - a parametrization
that has a direct and obvious physical meaning.
Data mining only provides a parametrization effectively isomorphic to the one by $p$:
to the eye the $p$-$v_2$ function appears continuous and with a continuous inverse.
Providing a physical meaning for the parameterization discovered (or finding a physically
meaningful parameterization isomorphic to the one discovered) is a completely separate
task, where the modeler has to provide good candidates.
The contribution of the data-mining process is determining
the {\em number} of necessary parameters, and in providing a quantity against
which good candidates can be tested.

\subsection{\label{ss:cl} Test case $2$: A two parameter family of graphs}

\begin{figure}
\begin{center}
\includegraphics[width=0.81\textwidth]{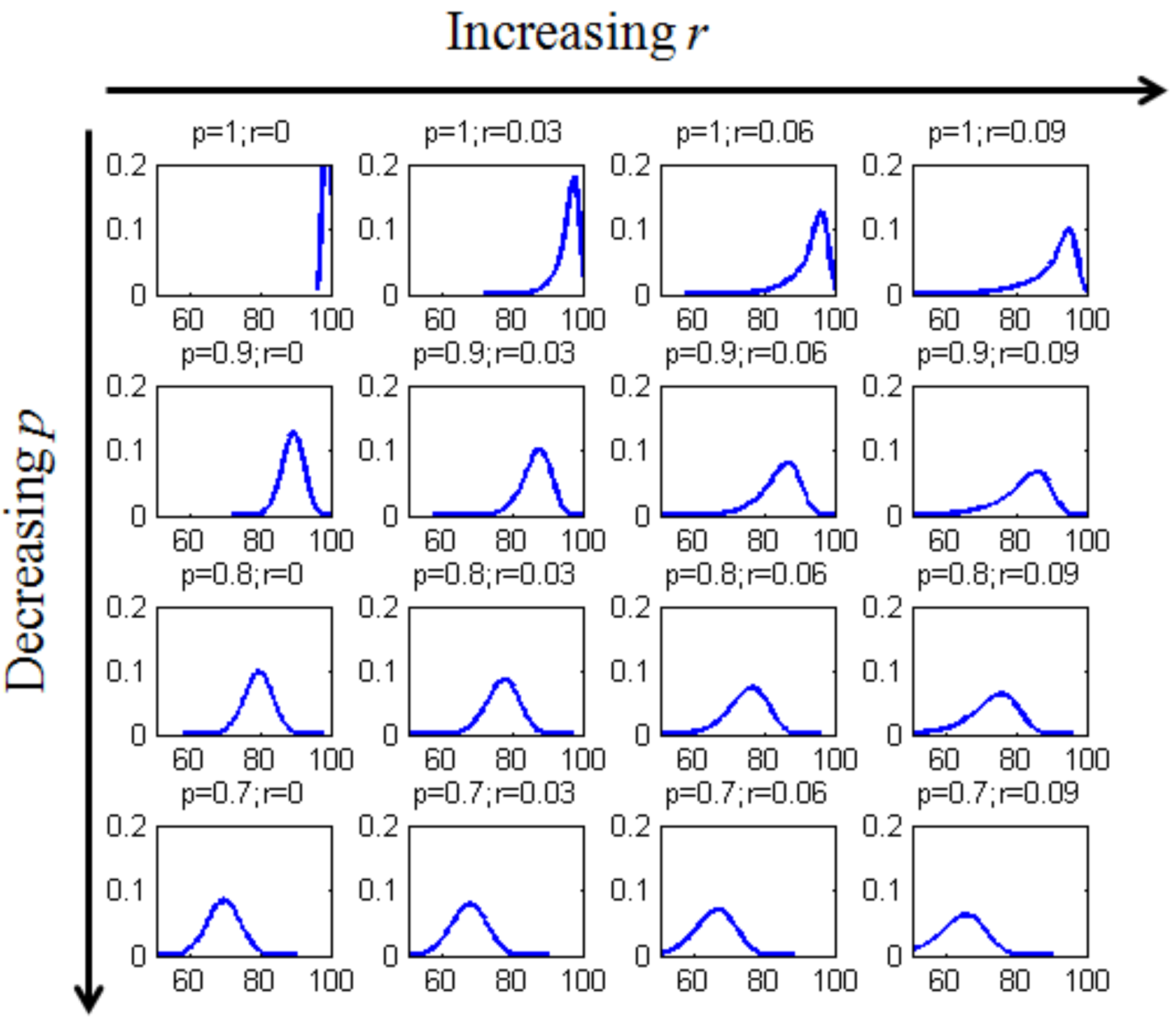}
\caption{\label{fig:CL} The degree distribution of Chung-Lu graphs created using
the algorithm described in the text are plotted for various values of the construction parameters
$p$ and $r$. The parameter $p$ corresponds to the density of edges in the
graph. As $p$ decreases, the degree distribution shifts uniformly to the left.
The parameter $r$ corresponds roughly to the skewness of the degree distribution.
As $r$ is increased from $0$, the degree distribution shifts to the left, but the
resulting degree distributions are skewed more and more to the left.
}
\end{center}
\end{figure}

\begin{figure}
\begin{center}
\includegraphics[width=0.81\textwidth]{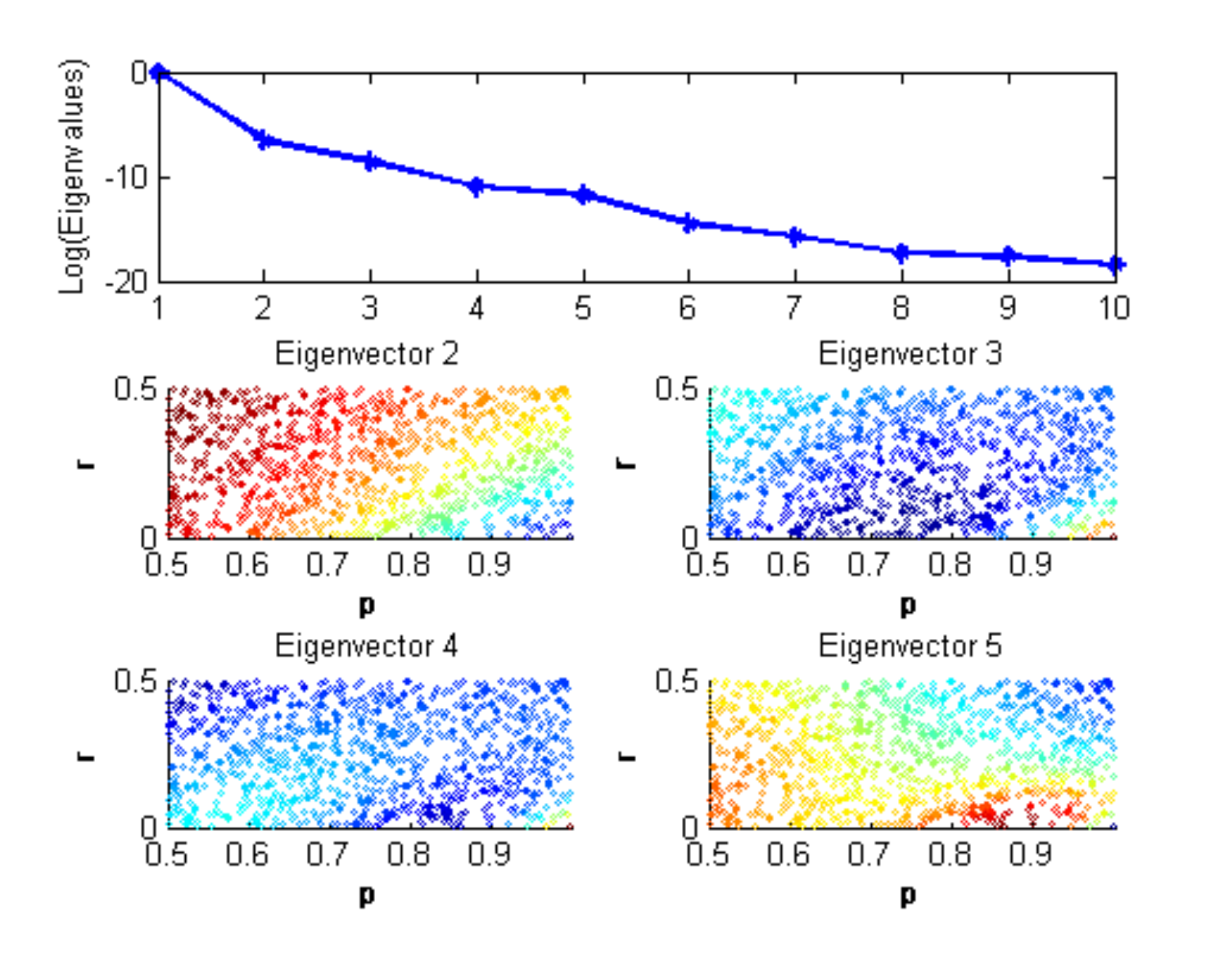}
\caption{\label{fig:CL1} Data mining ensembles of two-parameter Chung-Lu graphs:
The leading eigenvalues of the random walk matrix calculated using the subgraph similarity measure are first plotted.
The corresponding first four non-trivial eigenvectors are then illustrated in a way that brings forth their
relation to the construction parameters $p$ and $r$.
In these plots, each graph is denoted as a point.
The $x$ and $y$ coordinates of the point correspond to the parameters $p$ and $r$ used to construct that particular graph.
The graphs are colored based on the magnitude of their components in the eigenvectors of the random walk matrix $A$.
}
\end{center}
\end{figure}

\begin{figure}
\begin{center}
\includegraphics[width=0.81\textwidth]{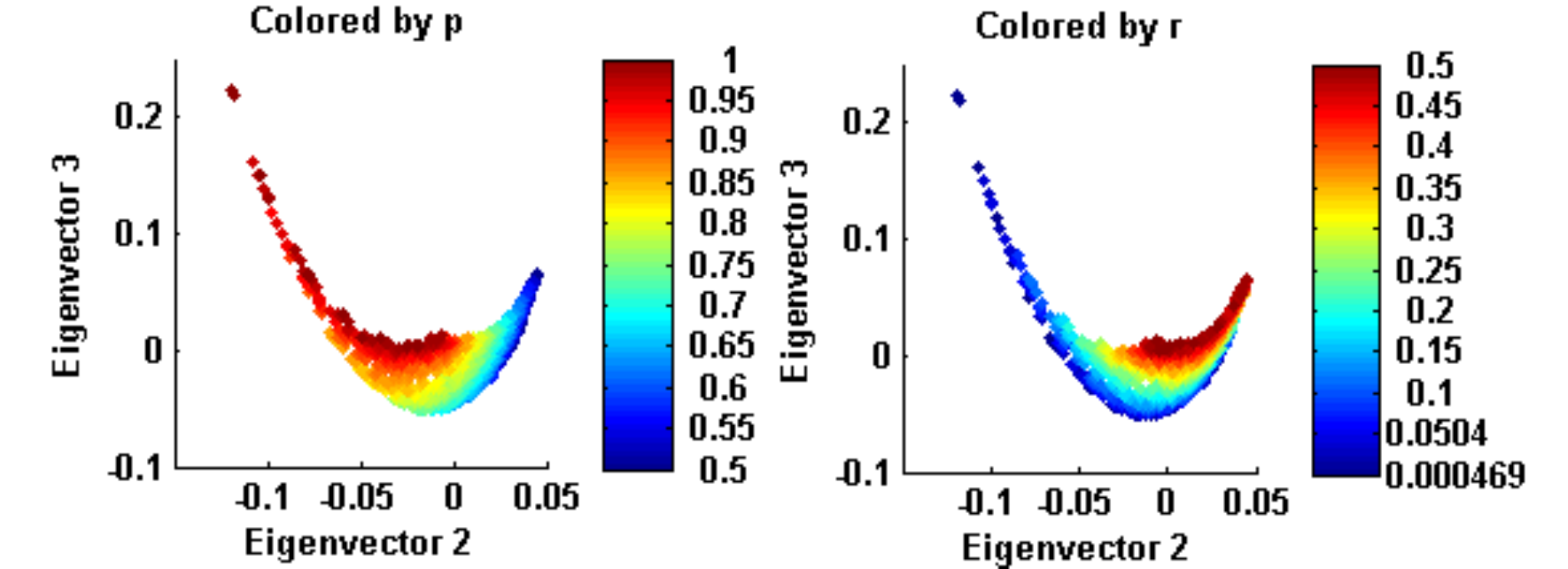}
\caption{\label{fig:CL1a}
Data mining the two-parameter family of Chung-Lu graphs using the subgraph similarity metric
leads to an apparent two-dimensional
embedding.
In these plots the $x$ and $y$ coordinates of each point (i.e. of each graph in the dataset)
denote the components
of that particular graph in the second and third eigenvectors of the random walk matrix respectively.
Each point is now colored
based on the parameter values of $p$ (left) and $r$ (right) used to construct the particular graph.
}
\end{center}
\end{figure}

\begin{figure}
\begin{center}
\includegraphics[width=0.81\textwidth]{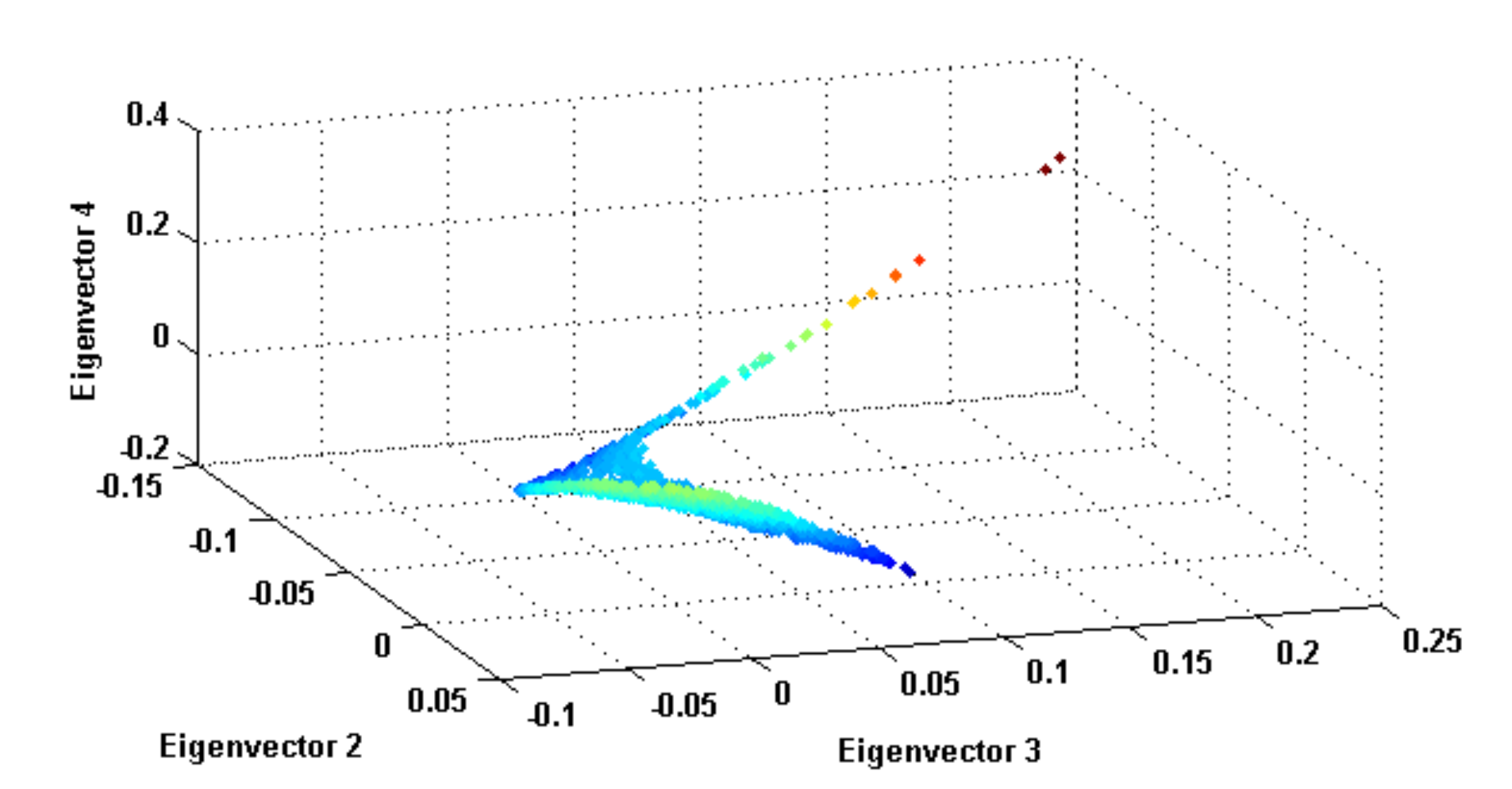}
\caption{\label{fig:CL1b} A $3-d$ plot suggesting that the fourth eigenvector of
the random walk matrix -calculated using the subgraph similarity measure- for the case of the two parameter family of
Chung-Lu graphs
can be expressed as a function of the second and third
eigenvectors: it does not capture a new direction in the space of our sample graphs.
}
\end{center}
\end{figure}

\begin{figure}
\begin{center}
\includegraphics[width=0.81\textwidth]{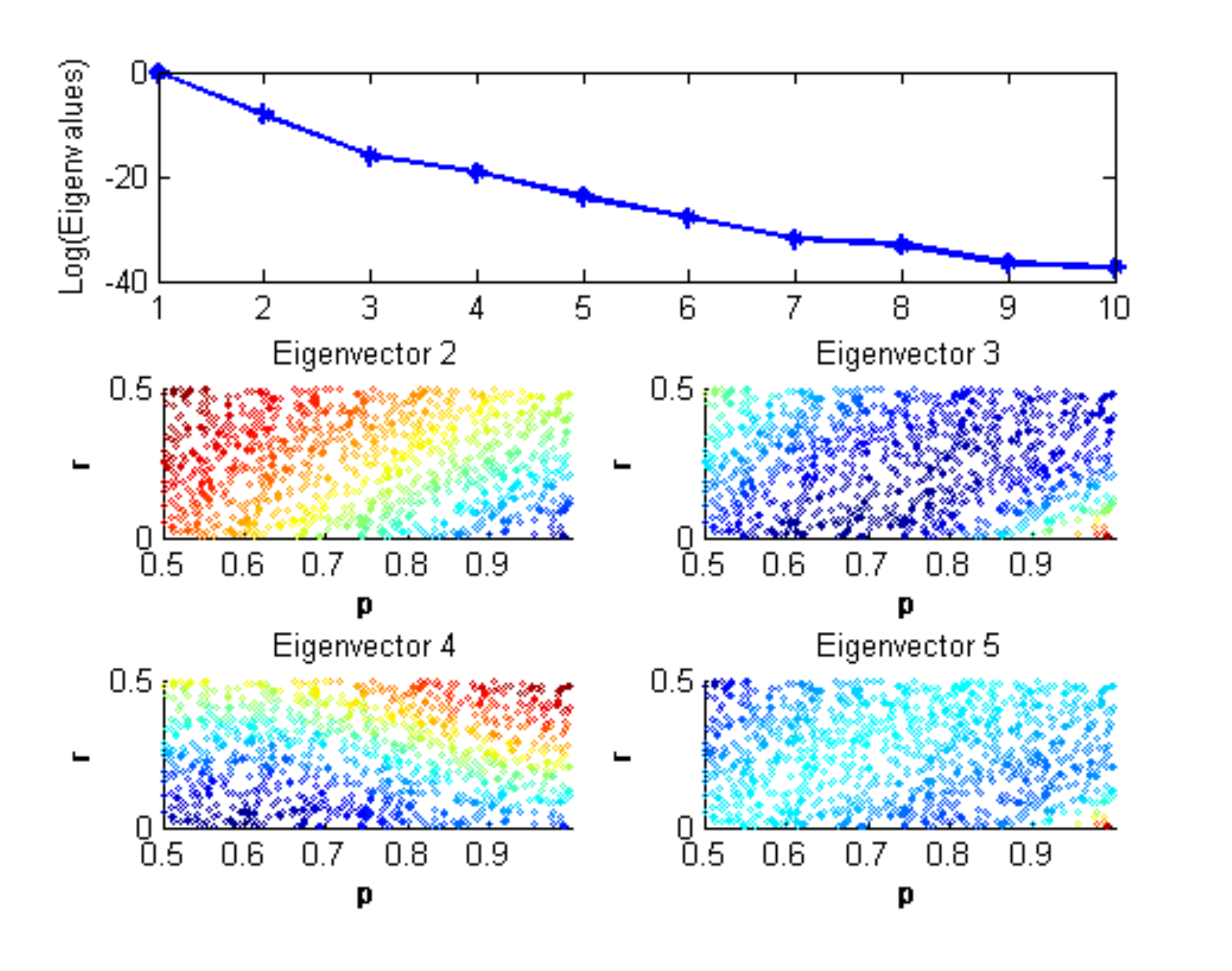}
\caption{\label{fig:CL2} Data mining ensembles of two-parameter Chung-Lu graphs:
The leading eigenvalues of the random walk matrix calculated using our spectral similarity measure are first plotted.
The corresponding first four non-trivial eigenvectors are then illustrated in a way that brings forth their
relation to the construction parameters $p$ and $r$.
In these plots, each graph is denoted as a point.
The $x$ and $y$ coordinates of the point correspond to the parameters $p$ and $r$ used to construct that particular graph.
The graphs are colored based on the magnitude of their components in the eigenvectors of the random walk matrix $A$.
}
\end{center}
\end{figure}

\begin{figure}
\begin{center}
\includegraphics[width=0.81\textwidth]{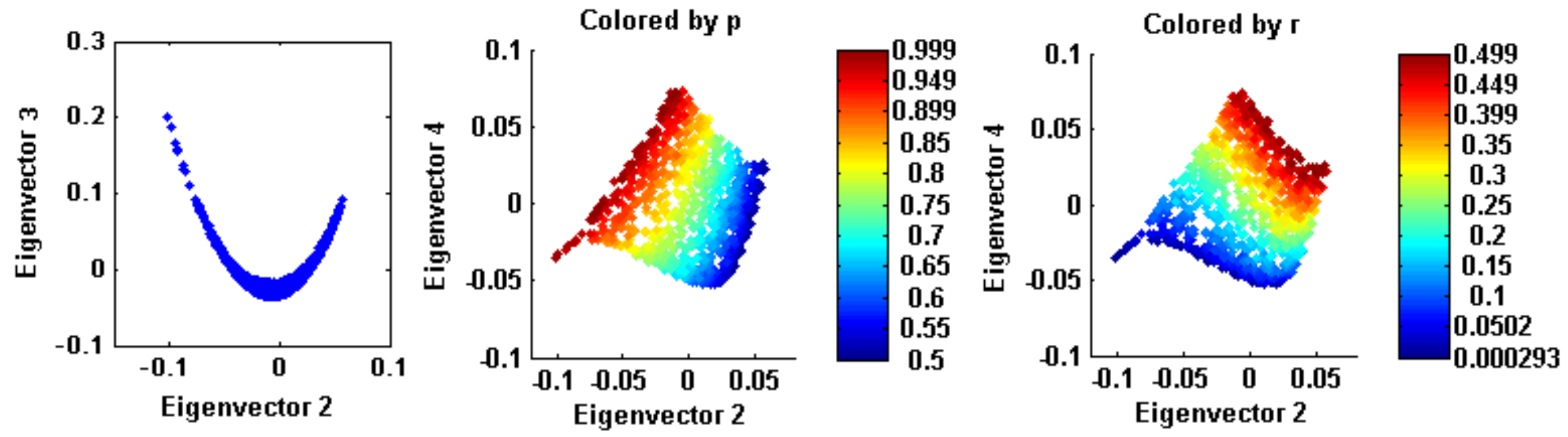}
\caption{\label{fig:CL2ab} Data mining the two-parameter family of Chung-Lu graphs using the subgraph similarity metric
leads to an apparent two-dimensional
embedding.
In these plots the $x$ and $y$ coordinates of each point (i.e. of each graph in the dataset)
denote the components of that particular graph in the second and third (resp., fourth) eigenvectors of the random walk matrix 
for the left plot (resp., middle and right plots).
Each point is also colored based on the parameter values of 
$p$ (middle plot) and $r$ (right plot) used to construct the particular graph.
}
\end{center}
\end{figure}

We now consider a slightly richer dataset, where
the graphs are constructed using two independent parameters.
The definition of this illustrative family of graphs is based
on the Chung-Lu algorithm \cite{Chun02connected}.
For a graph consisting of $n$ vertices (here $n=100$),
following their original algorithm, we begin by assigning
a weight $w_i$ to  each vertex $i, 1 \leq i \leq n$
The weights we chose have the two-parameter form  $w_i = np(i/n)^r$.
The probability $P_{ij}$ of existence of the edge between vertices
$i$ and $j$ is given by $P_{ij}=min(Q_{ij},1)$, where

\begin{equation}\label{}
Q_{ij} = \frac{w_iw_j}{\sum_{k}{w_k}}.
\end{equation}

Once the edge existence probabilities are calculated,
a graph can be constructed by sampling uniform random
numbers between $0$ and $1$ for every pair of vertices $(i,j)$
and placing an edge between them if the random number
is less than $P_{ij}$.
Note that in the original Chung-Lu algorithm $P_{ij}=Q_{ij}$.
If the weights are chosen such that $Q_{ij} <= 1, \forall (i,j)$,
then the expected value of the degree of node $i$ would
be equal to the chosen weight values $w_i$.
If any $Q_{ij}$ exceeds the value of $1$, this would no longer be the case \cite{Chun02connected}.

The model selected here has $2$ construction parameters: $p$ and $r$.
If $r=0$, the resulting graphs are Erd\"{o}s-R\'{e}nyi
graphs and the parameter $p$ represents the edge density.
When $p=1$ and $r=0$, the resulting graphs are complete.
As $r$ is increased, this procedure creates graphs whose
degree distributions are skewed to the left (long tails towards lower degrees).
The degree distributions resulting from creating graphs with various combinations
of parameters $p$ and $r$ are shown in Fig.~\ref{fig:CL}.

For our illustration, $1000$ graphs were created using
this model with $n=100$ nodes each.
The values of $p$ and $r$ were chosen by uniformly
sampling in the interval $(0.5,1)$ and $(0,0.5)$ respectively.
The diffusion maps algorithm was used on this set of graphs
exactly as described in the first case.
As we will discuss below, the results obtained using the
two similarity measures that we consider in this paper,
while conveying essentially the same qualitative information,
have visible quantitative differences.

The first $10$ eigenvalues of the random walk matrix calculated by
using the subgraph approach for evaluating similarities are shown in
the top plot of Fig.~\ref{fig:CL1}.
The first four non-trivial eigenvectors are plotted below.
In these plots, each of the $1000$ graphs is represented as a point
in the $p-r$ two parameter plane.
The colors represent the magnitude of the components of the corresponding
graph data on each of the first $4$ non-trivial eigenvectors.
The gradient of colors in these plots suggest the direction
of each of these eigenvectors in the $p-r$ plane.
However, a more careful inspection of the plots is required to determine
independent subsets of these eigenvectors. A quantitative approach to this issue can be found in \cite{dsilva2015parsimonious}.
To help explore this, we plot eigenvectors $2$ and $3$ against
each other in Fig.~\ref{fig:CL1a}.
The figure clearly suggests (through its obvious two-dimensionality) that these two eigenvectors are independent
of each other.
Furthermore, when the points in these plots are colored by the two parameters
$p$ and $r$ used to construct the graphs, two independent directions
- a roughly ``left-to-right" for $p$ and a roughly ``top-to-bottom" for $r$-
can be discerned on the $v_2 - v_3$ manifold, Fig.~\ref{fig:CL1a}.
This strongly suggests that the Jacobian of the transformation from $(p,r)$ to $(v_2,v_3)$
is nonsingular on our data.
Thus, these two eigenvectors, obtained solely through our data mining approach,
can equivalently be used to parameterize the set of graphs constructed using the parameters $p$ and $r$.
The components of the fourth eigenvector plotted in
terms of these two leading eigenvectors in Fig.~\ref{fig:CL1b}
are strong evidence that this fourth eigenvector is completely determined by
(is a function of) the second and third ones.
In other words, the fourth eigenvector ``lives in the manifold" created
by the second and third eigenvectors, and hence does not convey
more information about (does not span new directions in) our graph dataset.

We now focus on similar results obtained with the same dataset,
but now using our spectral approach for measuring similarity.
The eigenvalues and eigenvectors of the random walk matrix
obtained by this approach are reported in Fig.~\ref{fig:CL2}.
As before, we plot the leading eigenvectors against each other
in Fig.~\ref{fig:CL2ab}.
The plot of eigenvector $2$ versus eigenvector $3$ appears as a smooth ``almost" curve,
suggesting a strong correlation,
while the plot of eigenvector $3$ versus eigenvector $4$ clearly shows two-dimensionality.
These figures suggest that eigenvectors
$2$ and $3$ parameterize the same direction in
the $p-r$ plane, while eigenvector $4$ parameterizes a second, new
direction in this plane.
Hence, eigenvectors $2$ and $4$ constitute independent directions in the space
of our sample graphs.

Once again, we  have recovered (through data mining) two independent directions in our
sample family of graphs that were constructed using two independent parameters.
Although the results obtained using the subgraph and the spectral approaches
in this case are quantitatively different in their details, they are both successful in recovering
two independent coordinates in the space of graphs given as input to the data mining algorithm.
The $2D$ manifold resulting from the subgraph approach seems visually
better at visually capturing the behavior of the original $p-r$
plane.
The ``quality" of these parametrizations will clearly be affected by the
details used in the data-mining procedure and, in particular, those
affecting the similarity measure evaluation: the number of subgraph
densities kept, the choices for numerical constants such as
$\lambda$ in the subgraph approach, $\epsilon$ in diffusion maps, etc.
An obvious criterion in the selection of these method parameters is to make
the Jacobian of the transformation from the ``natural" to the ``data-mining-based"
parametrizations as far from singular as possible.

\subsection{\label{ss:dm} Test case $3$: Graphs from a dynamic graph evolution model}

In the two examples given above, the graphs in each dataset were created
using prescribed rules of a model controlled by one or more parameters.
Let us now consider a case where the graph dataset comes from sampling
graphs from a dynamical process at regular time intervals.
The dynamical process could either be an empirically evolving graph system
or a dynamic model with prescribed rules of evolution.
We consider the latter case for illustration, making use of a simple
model of a random evolution of networks \cite{bold2014equation}.
A brief description of the model is given here.
Starting from an initial graph, the model rules update the
graph structure at every time step by repeatedly applying the
following operations:

\begin{enumerate}
\item A pair of nodes selected at random are connected by an edge if they
are not already connected to each other.
\item An edge chosen uniformly at random is removed with probability $r=0.1$.
\end{enumerate}

\begin{figure}
\begin{center}
\includegraphics[width=0.81\textwidth]{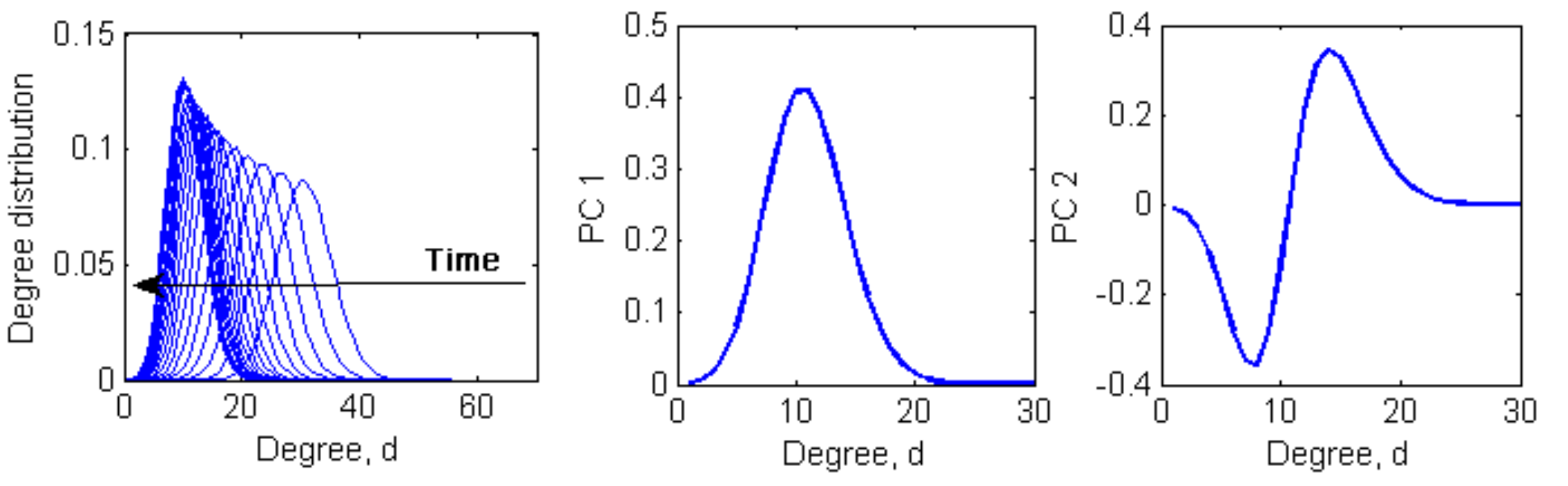}
\caption{\label{fig:K_d3} The evolution of degree distribution over
time from a single initial condition is shown on the left;
The first two principal components (obtained through PCA) of sequences
of degree distribution starting from different initial conditions
is shown on the right.
}
\end{center}
\end{figure}

The details of the model behavior are discussed in \cite{bold2014equation}.
Here, we will focus only on the characteristics of the model evolution
necessary to explain the results of data mining.
For this particular graphical model, it is known that the degree distribution
evolves smoothly in time as shown in the left plot of Fig.~\ref{fig:K_d3}.
Furthermore, it is also known that the evolution of degrees is decoupled from
(and slower than) the evolution of all higher order properties such as triangles,
degree-degree correlations, and so on.
Thus the information about the dynamic evolution of the graphs according to this
model can be sufficiently captured by studying the degrees.
A principal component analysis of sequences of degrees starting from different
initial conditions was performed, and the two leading principal components
are also plotted in Fig.~\ref{fig:K_d3}.
The first principal component (PCA), labeled PC $1$, corresponds the steady state degree distribution
while the second principal component PC $2$ corresponds to the direction along which
the degree distribution decays the slowest towards steady state.
Since the degree distribution is known to be the most significant variable in
this model, PC $1$ and PC $2$ are good variables to track the evolution of
the graphs over time.
In fact, as shown in  \cite{bold2014equation}, one can write explicit
Fokker-Planck equations for the evolution of the distribution of
(appropriately shifted and scaled) degrees.
The eigenfunctions of the corresponding eigenvalue problem are Hermite polynomials,
the first two of which have the qualitative forms, $exp(-x^2)$ and $x.exp(-x^2)$,
which are expressed in PC $1$ and PC $2$ respectively.

\begin{figure}
\begin{center}
\includegraphics[width=0.81\textwidth]{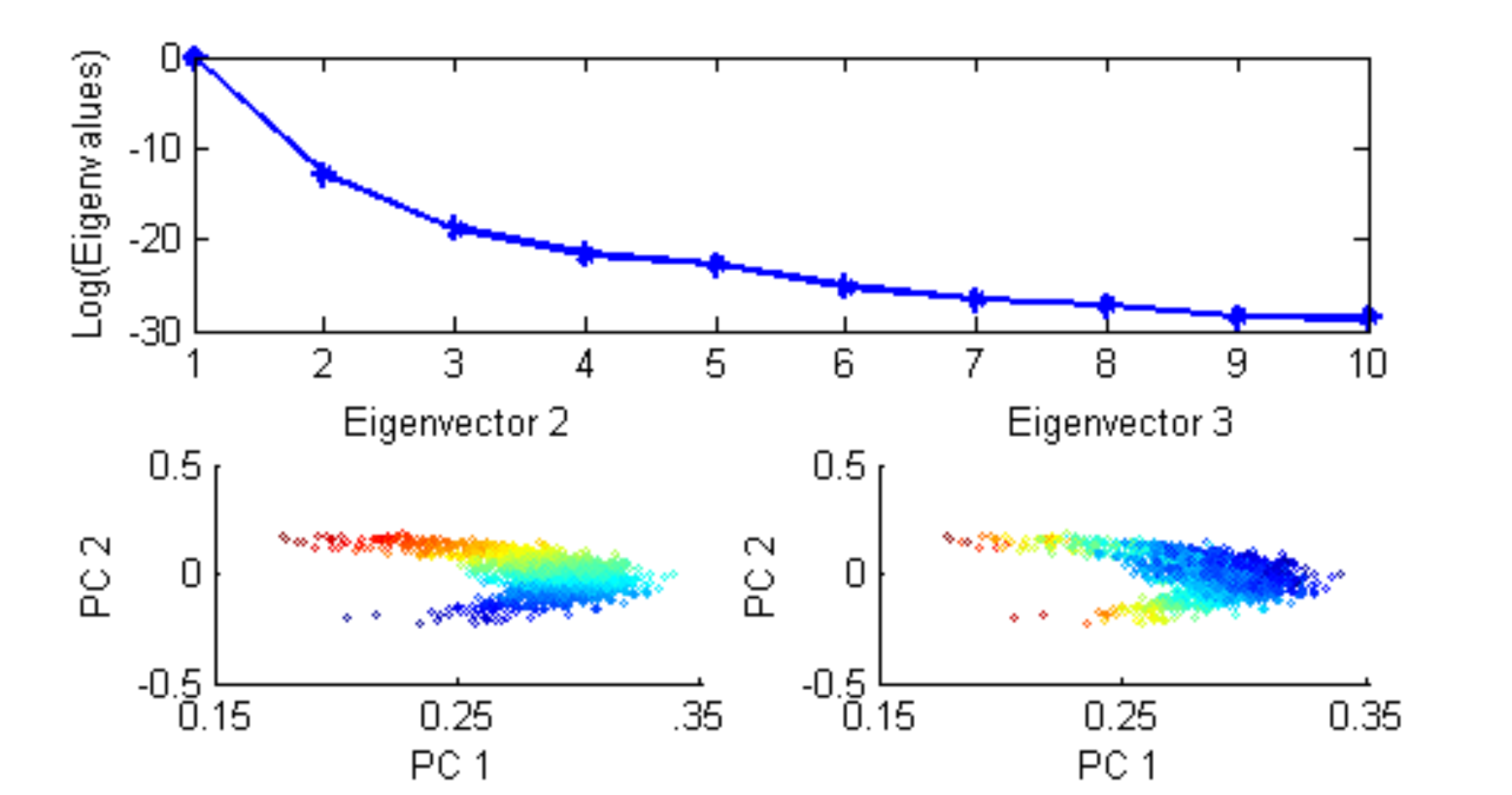}
\caption{\label{fig:K1} Data mining results for graphs collected from the dynamic graph evolution model:
The principal eigenvalues of the random walk matrix calculated using the subgraph similarity measure are plotted.
The corresponding first four non-trivial eigenvectors are shown here. In these plots, each graph is denoted as a point.
The $x$ and $y$ coordinates of the point correspond to the parameters PC $1$ and PC $2$ described in the text.
The graphs are colored based on the magnitude of the eigenvectors of the random walk matrix $A$.
}
\end{center}
\end{figure}

\begin{figure}
\begin{center}
\includegraphics[width=0.81\textwidth]{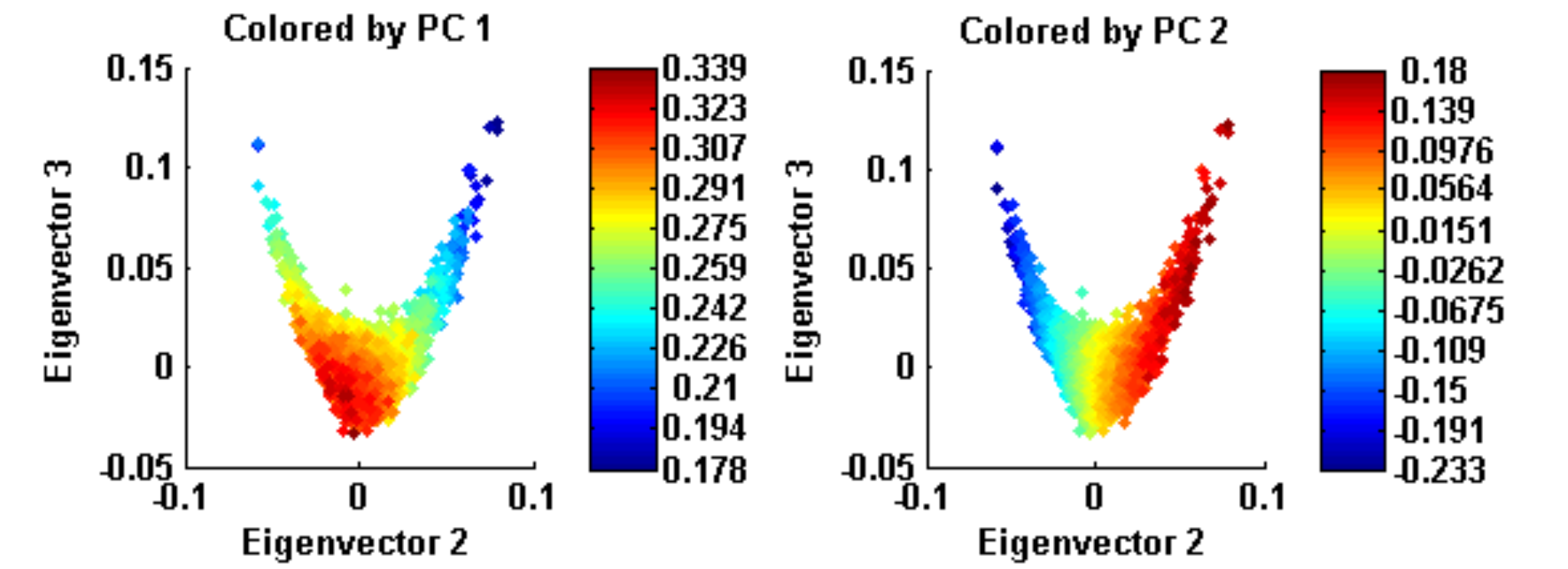}
\caption{\label{fig:K1a} The graphs collected from the dynamic model are plotted
in terms of the two eigenvectors shown in Fig.~\ref{fig:K1}. The points corresponding
to the different graphs are colored based on the first two principal components of the degree
distribution corresponding to these graphs.
}
\end{center}
\end{figure}

\begin{figure}
\begin{center}
\includegraphics[width=0.81\textwidth]{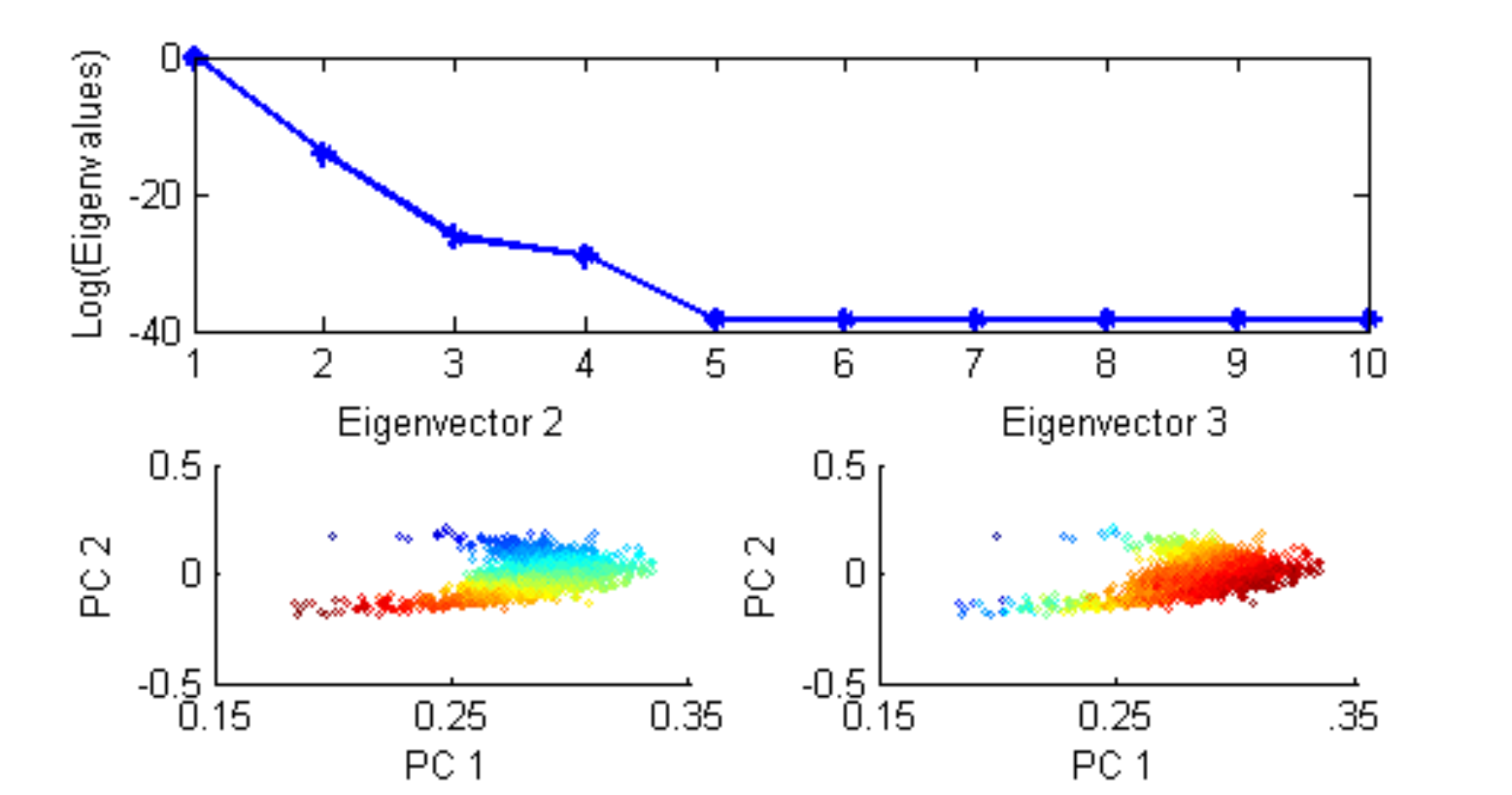}
\caption{\label{fig:K2} Data mining results for graphs collected from the dynamic graph evolution model:
The principal eigenvalues of the random walk matrix calculated using the spectral similarity measure are plotted.
The corresponding first four non-trivial eigenvectors are shown here. In these plots, each graph is denoted as a point.
The $x$ and $y$ coordinates of the point correspond to the parameters PC $1$ and PC $2$ described in the text.
The graphs are colored based on the magnitude of the eigenvectors of the random walk matrix $A$.
}
\end{center}
\end{figure}

\begin{figure}
\begin{center}
\includegraphics[width=0.81\textwidth]{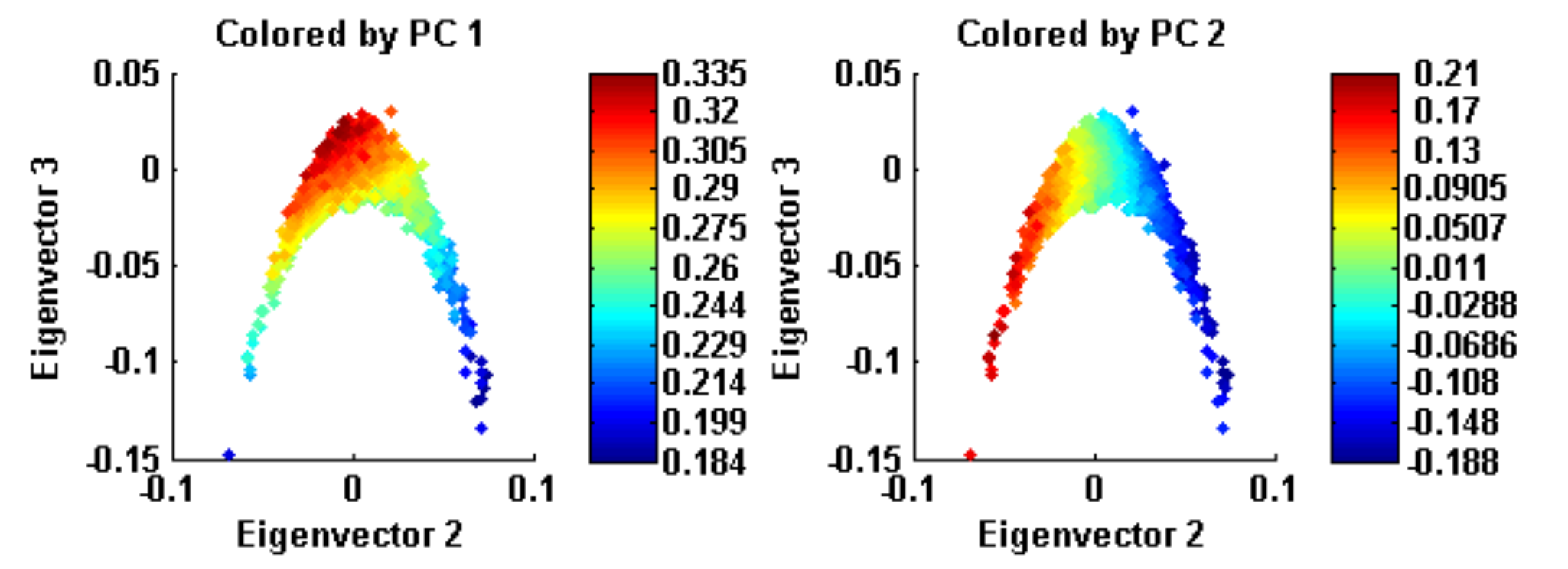}
\caption{\label{fig:K2a} The graphs collected from the dynamic model are plotted
in terms of the two eigenvectors shown in Fig.~\ref{fig:K2}. The points corresponding
to the different graphs are colored based on the first two principal components of the degree
distribution corresponding to these graphs.
}
\end{center}
\end{figure}

Let us now ignore our knowledge of the dynamical process and only consider the data: the graph sequences created by the process.
Our goal is to use diffusion maps to find out good variables to characterize
these graphs.
Since the graph data used for data mining come from a dynamical process, the
variables obtained through data mining should correspond to the dynamics of the
process.
In other words, we expect the data mining variables and the variables PC $1$ and
PC $2$ to convey similar information about the graphs.
As before, we will use both the subgraph and spectral similarity measures to
get the results.
The eigenvalues and the first two non-trivial eigenvectors of the random walk
matrix in diffusion maps are shown for both the subgraph and spectral similarity
measures in Figs.~\ref{fig:K1} and \ref{fig:K2} respectively.
The plots are shown such that the points (corresponding to the graphs)
are plotted in the plane of the principal components PC $1$
and PC $2$ and are colored by the eigenvectors for comparison.
In both cases, the results show that the second and third eigenvectors have the most variation
(as indicated by the gradient of colors) in the directions of PC $2$ and PC $1$ respectively.
Conversely, we can plot the graphs in terms of the diffusion map eigenvectors (the new embedding)
and color them based on PC $1$ and PC $2$ as shown in Figs.~\ref{fig:K1a} and \ref{fig:K2a}
corresponding to the subgraph and spectral similarity measures respectively.
This indicates that the first two non-trivial eigenvectors are roughly one-to-one with
PC $2$ and PC $1$ respectively.

\begin{figure}
\begin{center}
\includegraphics[width=1\textwidth]{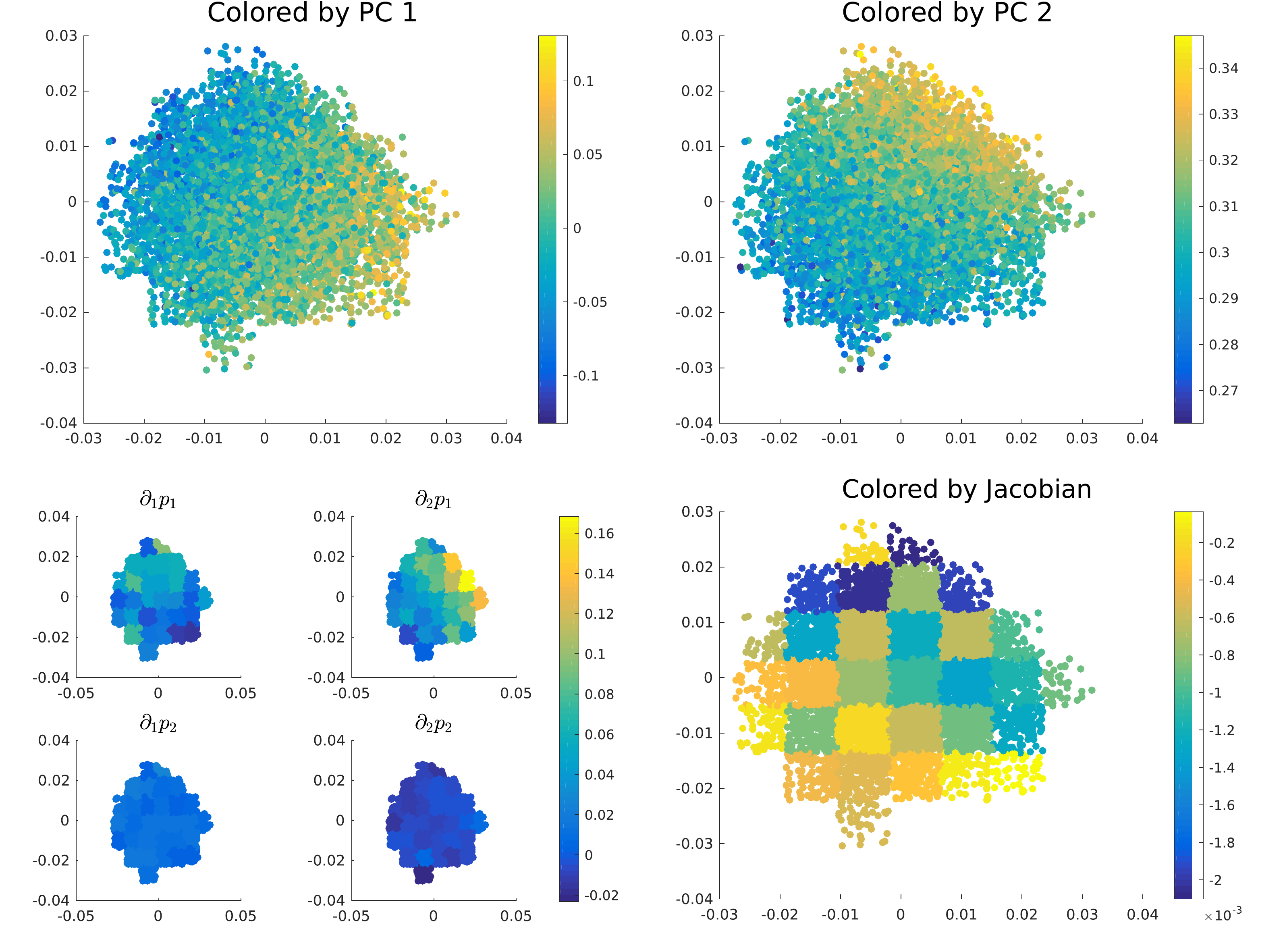}
\caption{\label{fig:J} Data mining results for graphs collected from the dynamic graph evolution model for a large dataset are shown above, colored by PC $1$ and PC $2$ respectively. The two eigenvectors deemed significant here were the second and fourth, with the same DMAP procedure performed as before. Below, each graph datum is colored by the individual partial derivatives on the left and by the Jacobian on the right. We easily notice that $\partial_{\phi_1} p_2$ remains positive everywhere, as was expected, while the Jacobian does not change sign - indicating that the transformation is one-to-one.
}
\end{center}
\end{figure}

For a quantitative verification of this observation, we can consider the mapping $f : \boldsymbol{\phi} \rightarrow \boldsymbol{p}$ between the diffusion map eigenvectors $ \boldsymbol{\phi} = (\phi_2, \phi_3)$ and the two principal components $\boldsymbol{p} = (p_1, p_2)$. By suitably discretizing the space of graph snapshots, we can directly compute the average rate of change of each principal component with respect to each eigenvector coordinate. This is achieved by considering all graphs in a specific neighborhood of the diffusion map space and averaging the rate of change of $p_{\{1,2\}}$ with respect to each of $\phi_1$ and $\phi_2$ within that neighborhood, thus providing a local estimate of each partial derivative $\partial_{\phi_j} p_i $. As outlined in Fig. \ref{fig:J}, this allows us to verify that the Jacobian of $f$ is non-zero everywhere, giving further evidence that the transformation between the two is bounded away from zero.

To summarize, we have shown an illustrative example here in which
graphs collected from a dynamic process were used to recover important variables
parametrizing the evolution of the process.
We considered a simple example for which theoretical results were available,
so that we were able to compare the results from data mining with those from theory.
In problems where such theoretical results are not available, one can use
data mining  to gain an understanding about the primary driving factors
in the dynamics of the system.

\section{\label{sec: ef} Equation-Free Methods for Accelerating Graph Evolution Computations}

By extracting the important variables governing the evolution of a graph-based dynamical process, we can investigate the mapping between these variables and temporal instances of the process. To do so, we turn our attention to Equation-Free (EF) modeling methods \cite{kevrekidis2003equation} \cite{kevrekidis2004equation} and their applications to network-based dynamical systems. EF methods provide a framework for working with low-dimensional representations (``coarse variables") of a dynamical system, even where closed form expressions of these variables are not available. More specifically, we proceed in two steps: firstly, a `coarse-graining' technique is used to extract a meaningful parametrization of the dynamics of the process in question, i.e. one encompassing all of its degrees of freedom in a lower dimensional representation. If such a low-dimensional characterization of the system exists, then this step identifies both its inherent dimensionality and the suitable coarse variables that span it. Here, we use the variables recovered by our data-mining approach as the required coarse variables that parametrize it. The second step involves the investigation of mappings between these coarse variables and instances of the dynamical process. We are thus required to construct suitable mapping operations between the coarse variables identified in the previous step and instances of the dynamical process in an efficient fashion. This is discussed in more detail below.

\subsection{Lifting \& Restriction}

The two key ingredients for the use of EF techniques are the capability to move from instances of the `fine-grained' system (in this case realizations of the evolving network) to the `coarse-grained' representation of the system and vice-versa. We denote the operators defining the movement between these two regimes as {\em restriction} and {\em lifting} respectively, with the latter usually posing a much larger computational challenge and involving multiple instances of the former. More specifically, if we denote by $\psi_t$ the fine-grained temporal evolution operator of the dynamical process, we can define the coarse-grained evolution operator $\Psi_t$ as follows:
\begin{equation}
\Psi_t ( \cdot ) = R \circ \psi_t \circ L ( \cdot ),
\end{equation}
\noindent where $R(\cdot)$ and $L(\cdot)$ denote the restriction and lifting operations respectively. 

The main idea underpinning this method is that, given a suitable coarse description of the system and efficient lifting/restriction operators, we do not have to work exclusively in the fine-grained regime (i.e. by executing repeated long-term system simulations), which is usually computationally expensive. Instead, we can restrict to the coarse variables, advance the system in this regime (usually using numerical estimation techniques), and then lift back to an instantiation of the fine-grained system that is suitably close to where direct simulation would have advanced the process. Indeed, with the help of these operators, appropriately initialized instances of the dynamical system are run for short bursts of time, providing local information that can then be used to expedite analysis of the system's dynamics.

\subsection{Coarse-Projective Integration}

Various methods for the implementation of EF techniques in network-based dynamical systems have been explored \cite{bold2014equation} \cite{holiday2016equation}, but they usually require an intimate understanding of the dynamical system in question in order to implement the coarse-graining step. As mentioned previously, here we will use the data mining procedure from Section~\ref{sec:dm} to coarse-grain the dynamical process, without making any explicit assumptions about its dynamics. This has the advantage of requiring no previous information about the system in question, and thus can be applied in a very general context. We illustrate the confluence of the proposed data mining technique with EF methods by demonstrating its application to Coarse-Projective Integration (CPI) \cite{gear2003projective}, an EF technique whose primary goal is the acceleration of the dynamical system's time evolution by projectively integrating the coarse variables forward in time. In a comparative analysis, we demonstrate that the temporal evolution of the underlying dynamical system can be accelerated through CPI. 

To illustrate this, we work with the dynamical network example from the previous section and compare our results of both a long-term fine-grained simulation and the CPI simulation. We furthermore look at the underlying variables, which are known to define long-term dynamics, of these two simulations at different timesteps. This allows us to further validate our approach. In the context of EF computations, we use the two leading eigenvectors $(\phi_2, \phi_3)$ as the coarse-graining variables of interest, which are obtained through our data-mining procedure. More specifically, we begin by generating a large (with $\sim 10^3$ datapoints) `reference' graph dataset that contains snapshots of networks approaching the system steady state from many directions. This is done by taking temporal snapshots of many Erd\"{o}s-R\'{e}nyi random graphs evolving according to the network rules, exactly as in the preceding section, and performing DMAPs on the ensemble to get the eigenvector coordinates of each graph datum. For information and details on how we defined the restriction and lifting operators, we refer the reader to the Appendix. 

\begin{figure}
\begin{center}
\includegraphics[width=0.81\textwidth]{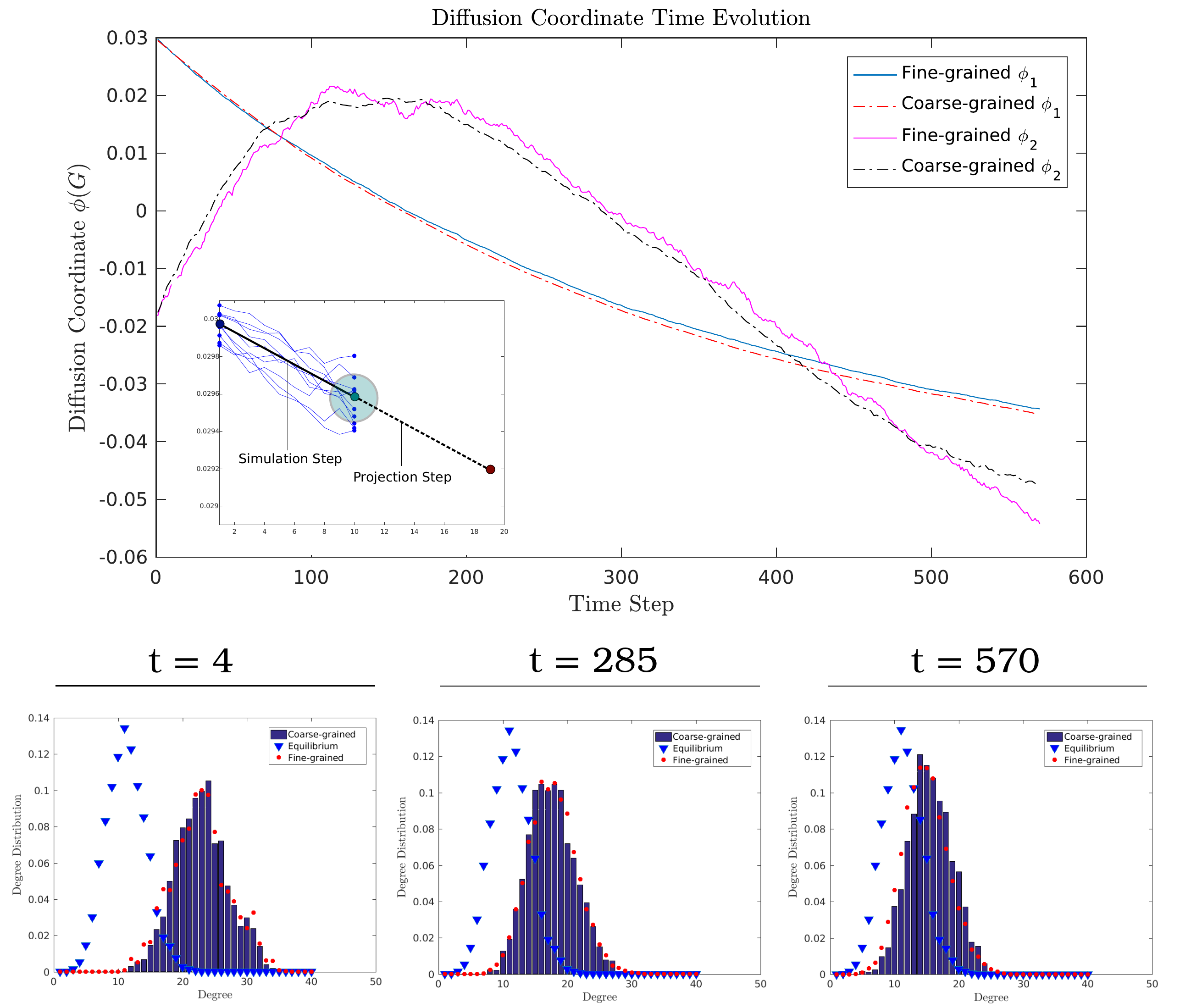}
\caption{\label{fig:CPI} The two diffusion coordinates of the network $G$ after each timestep are shown above for both the CPI and fine-grained simulations, with both beginning from the same initial graph. Below, the degree distribution from CPI and the fine-grained simulation are shown alongside the equilibrium distribution for reference. It should be noted that the `drift' towards the equilibrium distribution over time is captured both by the CPI (blue bars) and fine-grained (red) temporal evolution. In the inset, one step in the CPI process is illustrated, with the simulation and projection steps over time of the first eigenvector from $t = 0$ shown for clarity. Each timestep here denotes $10$ iterations of the rules of the process, each short `burst' of simulation lasts for $t_B = 10$ timesteps, and we project the coarse variables forward in the CPI step by another $t_P = 10$ timesteps, effectively halving the total number of steps required. The subgraph metric with $\epsilon = 10$ was used in generating the reference data.
}
\end{center}
\end{figure}

To implement CPI, we simulate the system for a short burst of time $t_B$, keeping track of the diffusion coordinates of the underlying network before and after the simulation. By averaging over $k$ such short runs, we can then projectively integrate the diffusion coordinates forward by $t_P$ steps at a time faster than is possible with the fine-grained simulation. This is achieved through the use of Euler's method, although many techniques would suffice here. After reaching the `projected' diffusion coordinates, this process is repeated, providing a more efficient method for the temporal evolution of the underlying system. It should be noted, however, that this technique is only possible given the existence of not only the coarse grained system representation, but also the lifting and restriction operators. In Fig.~\ref{fig:CPI}, we plot comparisons between the estimated diffusion coordinate values obtained over time for an instance of the dynamical system evolving through CPI and one evolving through fine-grained simulation. The close agreement between the two provides strong indications that CPI can be successfully used to aid temporal development of graph-based dynamical systems. It should also be noted that the known underlying coarse variable, the degree distribution, of this system shows very strong agreement in both the CPI and fine-grained runs. This is even stronger evidence that CPI not only shows small deviations from the actual simulation, but also that the important network properties underlying the system's long-term dynamics are captured by CPI.

\section{\label{sec:conc} Conclusions}

In this paper, we discussed the problem of data mining in cases where
the data points occur in the form of graphs.
The main obstacle to applying traditional data mining algorithms to
such cases is the definition of good measures to quantify the similarity
between graph pairs.
We discussed two common sense approaches to tackle this problem: {\em the subgraph
method}, which compares the local structures in the graphs, and {\em the spectral
method}, which is based on defining diffusion processes on the graphs.
While alternate definitions of similarity metrics than the ones discussed in
this paper are possible, the purpose of this paper was to demonstrate the
usefulness of data mining in the context of graphs, using a few illustrative
examples for which the parameterizations obtained through our approach
could be compared with known results.
Nevertheless, certain remarks need to be made regarding the similarity measures
used in this paper.
The subgraph approach to evaluate similarity is much more expensive compared to the
spectral approach especially when larger sized subgraphs are required to get
accurate results.
(For example, there are $6$ connected subgraphs of size $4$,
while there are $21$ subgraphs of size $5$. It is also computationally more
expensive to search for larger subgraphs).
Both approaches require us to tune certain parameters associated with  the
definitions of the similarity metric.
For the diffusion map algorithm, one has to choose a suitable size of
neighborhood ($\epsilon$).
In addition, the spectral approach required one to define the weighting function, $\mu(k)$
(and also make assumptions about the vectors $p$ and $q$).
This degree of freedom is roughly equivalent to selecting suitable normalizations
to find the subgraph densities in the subgraph approach.
These tuning considerations become especially crucial when one is confronted
with data from a fresh problem, where intuition cannot be used to guide
the selection of these parameters.
Considering the trade-offs mentioned above, it might be prudent to use
the subgraph density approach to find similarities between graphs initially for new problems
and tune the spectral decomposition algorithm, which can then be used for faster computations.

Having discussed the approach used for defining graph similarities and subsequently
data mining, let us now consider the problem from the point of view of applications.
We used three sample sets of graph data in this work.
The first example was a collection of Erd\"{o}s-R\'{e}nyi random graphs
with varying parameters.
We also considered the case of graphs obtained from a simple $2$ parameter
family of graphs motivated by the Chung-Lu algorithm.
Both these examples considered graphs created from a fixed model.
As a third example, we used a collection of graphs from a dynamic model.
In all these examples, we used the data mining approach with two different
approaches for measuring similarities to extract good characterizations of the
graph datasets and compared them to known parameterizations.
An obvious extension of the work in this paper
is to test the methods illustrated here
on datasets with graphs of varying sizes.
The similarity measures discussed in the paper were chosen so
that it is straightforward to extend them to such datasets.

Finally, the data mining technique based on these similarity measures was used on the data from the above dynamical system to speed up computations of its temporal evolution. Here we have made use of both the data mining procedure above and EF methods. Thus, this kind of result can be achieved with no need to fall back on theoretical knowledge about the process in question, which for many complex systems might not be available. Indeed, this coarse-level system description can, for example, then be used to swiftly advance the system through time and to perform an expedited analysis of the network's dynamics. We believe that the lifting and restriction operators defined here, coupled with the data mining technique above, can be used in dynamical systems based on networks for which we do not have closed form or `intuitive' expressions for the dynamics.

\section*{Appendix}
\addcontentsline{toc}{section}{Appendix}

In order to define efficient lifting and restriction operators, we work with a generated `reference' dataset $\{ G^i_R \}_{i = 1}^M$ of graph-diffusion variable pairs for a wide variety of network instances. We denote the diffusion variables of some graph $G_i$ in the reference dataset by $\phi_{ref}(G_i) = (\phi_2^i, \phi_3^i)$. It should be noted that throughout the simulation, we will always have access to both the network instances and corresponding eigenvectors of this precomputed dataset. Finally, it should be noted that we are going to be working not with individual graphs, but rather with {\em ensembles} of $N$ graphs in both transformations.

\subsection{Lifting}

We define the lifting operator as a mapping $L: \mathbb{R}^2 \rightarrow \{(c_i,G_i)\}^N_{i = 1}$, from one diffusion coordinate $\phi_0 = (\phi^0_2, \phi^0_3) \in \mathbb{R}^2$ to an ensemble of $N$ different graph-coordinate pairs, where $c_i \in \mathbb{R}$ the coefficient associated with graph $G_i$. We pick the ensemble of $N$ graphs from the reference dataset by looking at graphs whose diffusion coordinates are closest to $(\phi_2^0, \phi_3^0)$, i.e. the $N$ nearest neighbors by diffusion distance. To compute the corresponding coefficients $c_i$, we solve the following interpolation problem:

\begin{equation}
\label{eq:intra}
\phi_0 = \sum_{i = 1}^N c_i \cdot \phi_{ref}(G_i),
\end{equation}
which can be easily achieved through Singular Value Decomposition (SVD) techniques, since we choose $N > 2$. Note that the coefficients assigned to each graph denote the graph's `weight' in characterizing $\phi_0$ based on its own diffusion coordinates. In short, the lifting operator proceeds as follows:
\begin{enumerate}
\item On input $\phi_0$, find the $N$ reference graphs $\{G_i\}^N_{i = 1}$ whose diffusion coordinates $\{\phi_{ref}(G_i)\}_{i = 1}^N$ are closest to $\phi_0$.

\item For this collection of graphs, find the coefficients $c_i$ that solve \ref{eq:intra}. This is done by performing SVD on the linear system defined by \ref{eq:intra} and always admits a solution for $N > 2$.
\end{enumerate}

\subsection{Restriction}

We define the restriction operator $R: \{(c_i,G_i)\}^N_{i = 1} \rightarrow \mathbb{R}^2$ of some graph ensemble as the (approximate) diffusion coordinates of each $G_i$ weighted by their corresponding coefficients. This can be succinctly represented as:

\begin{equation}
R(\{(c_i,G_i)\}_1^N) = \sum_{i = 1}^N c_i \cdot \phi(G_i),
\end{equation}

\noindent where $\phi(G_i) \in \mathbb{R}^2$ is the approximate diffusion coordinate tuple of graph $G_i$. However, we should note here that the graphs being restricted might not be in the reference dataset, which means that we would need a way to calculate their diffusion coordinates. Instead of recomputing DMAPs every time, which would be computationally prohibitive, this is instead achieved through the use of Nystr\"om extension. This technique deals with the problem of finding the diffusion map coordinates of a {\em new} graph $G$ based on the already existing reference dataset. Although approximate, it suffices for our current purposes. 

The first step here is to calculate the new distances $\{d^i_{new}\}_{i = 1}^M$ between graph $G$ and each of the $M$ graphs in the reference dataset, using either the subgraph or spectral metrics. We then define $W^i_{new} = \exp{[-(d^i_{new}/\epsilon)^2 ]}$, where $\epsilon$ as in the reference data, and suitably normalize to yield:
\begin{equation}
K^i_{new} = \left( \sum_{k = 1}^M W_{new}^k \right)^{-1} W^i_{new}.
\end{equation}

\noindent We can then define the $j$-th diffusion map coordinate of graph $G$ as:
\begin{equation}
\phi_{new}(j) = \frac{1}{\lambda_j} \sum_{i = 1}^M K^i_{new} \cdot \phi_j(i),
\end{equation}

\noindent where $\phi_j(i)$ denotes the $i$-th coordinate of the $j$-th diffusion map eigenvector of the reference dataset and $\lambda_j$ the corresponding eigenvalue.

This allows us to `track' the development of the network in diffusion space by appealing only to a (pre-computed) reference dataset. Care must be taken, however, to include many network snapshots in the reference dataset that would be `close' in similarity to any network path we would want to model, as we will be using this dataset to approximate the coarse variables of networks that may look very different to each other. Ensuring that any fine-grained instantiation has sufficiently close `neighboring' reference graph snapshots in diffusion space (under Euclidean distance) substantially aids the accuracy of the lifting and projection mechanisms defined above.\\

\noindent \textbf{Acknowledgements}\\

\noindent The work of IGK was partially supported by the US National Science Foundation, as well as by AFOSR (Dr. Darema) and DARPA contract HR0011-16-C-0016.

\end{document}